\documentclass[twocolumn]{aastex62}

\usepackage{enumerate}

\received{...}
\revised{...}
\accepted{...}

\submitjournal{AJ}

\shorttitle{Multi-wavelength observations of the eccentric debris ring of HD 202628}
\shortauthors{Faramaz et al.}

\begin{document}

\title{From scattered-light to millimeter emission:\\ A comprehensive view of the Gyr-old system of HD 202628 and its eccentric debris ring}

\correspondingauthor{Virginie Faramaz}
\email{virginie.c.faramaz@jpl.nasa.gov}

\author{Virginie Faramaz}
\affiliation{Jet Propulsion Laboratory, California Institute of Technology, 4800 Oak Grove drive, Pasadena CA 91109, USA.}

\author{John Krist}
\affiliation{Jet Propulsion Laboratory, California Institute of Technology, 4800 Oak Grove drive, Pasadena CA 91109, USA.}

\author{Karl R. Stapelfeldt}
\affiliation{Jet Propulsion Laboratory, California Institute of Technology, 4800 Oak Grove drive, Pasadena CA 91109, USA.}

\author{Geoffrey Bryden}
\affiliation{Jet Propulsion Laboratory, California Institute of Technology, 4800 Oak Grove drive, Pasadena CA 91109, USA.}

\author{Eric E. Mamajek}
\affiliation{Jet Propulsion Laboratory, California Institute of Technology, 4800 Oak Grove drive, Pasadena CA 91109, USA.}

\author{Luca Matr\`a}
\affiliation{Harvard-Smithsonian Center for Astrophysics, 60 Garden Street, Cambridge, MA 02138, USA.}

\author{Mark Booth}
\affiliation{Astrophysikalisches Institut und Universit{\"a}tssternwarte, Friedrich-Schiller-Universit{\"a}t Jena, Schillerg{\"a}{\ss}chen 2-3, 07745 Jena, Germany.}

\author{Kevin Flaherty}
\affiliation{Department of Astronomy and Department of Physics, Williams College, Williamstown, MA 01267}

\author{Antonio S. Hales}
\affiliation{Joint ALMA Observatory, Alonso de C\'ordova 3107, Vitacura 763-0355, Santiago, Chile.}
\affiliation{National Radio Astronomy Observatory, 520 Edgemont Road, Charlottesville, Virginia, 22903-2475, USA.}

\author{A. Meredith Hughes}
\affiliation{Department of Astronomy, Van Vleck Observatory, Wesleyan University, Middletown, CT 06459, USA.}

\author{Amelia Bayo}
\affiliation{Instituto de F\'isica y Astronom\'ia, Facultad de Ciencias, Universidad de Valpara\'iso, Av. Gran Breta\~na 1111, Valpara\'iso, Chile.}
\affiliation{N\'ucleo Milenio Formaci\'on Planetaria - NPF, Universidad de Valpara\'iso, Av. Gran Breta\~na 1111, Valpara\'iso, Chile}

\author{Simon Casassus}
\affiliation{Departamento de Astronomia, Universidad de Chile, Casilla 36-D, Santiago, Chile.}
\affiliation{Millennium Nucleus "Protoplanetary Disks", Santiago, Chile.}

\author{Jorge Cuadra}
\affiliation{Instituto de Astrof\'isica, Facultad de F\'isica, Pontificia Universidad Cat\'olica de Chile, 782-0436 Santiago, Chile.}
\affiliation{N\'ucleo Milenio Formaci\'on Planetaria - NPF, Universidad de Valpara\'iso, Av. Gran Breta\~na 1111, Valpara\'iso, Chile}

\author{Johan Olofsson}
\affiliation{Instituto de F\'isica y Astronom\'ia, Facultad de Ciencias, Universidad de Valpara\'iso, Av. Gran Breta\~na 1111, Valpara\'iso, Chile.}
\affiliation{N\'ucleo Milenio Formaci\'on Planetaria - NPF, Universidad de Valpara\'iso, Av. Gran Breta\~na 1111, Valpara\'iso, Chile}

\author{Kate Y. L. Su}
\affiliation{Steward Observatory, University of Arizona, 933 N Cherry Ave., Tucson, AZ 85721, USA.}

\author{David J. Wilner}
\affiliation{Harvard-Smithsonian Center for Astrophysics, 60 Garden Street, Cambridge, MA 02138, USA.}

\begin{abstract}

We present here new observations of the eccentric debris ring surrounding the Gyr-old solar-type star HD 202628: at millimeter wavelengths with ALMA, at far-infrared wavelengths with \textit{Herschel}, and in scattered light with \textit{HST}. The ring inner edge is found to be consistent between ALMA and \textit{HST} data. As radiation pressure affects small grains seen in scattered-light, the ring appears broader at optical than at millimeter wavelengths. The best fit to the ring seen with ALMA has inner and outer edges at $143.1 \pm 1.7$ AU and $165.5 \pm 1.4$, respectively, and an inclination of $57.4^\circ \pm 0.4$ from face-on. The offset of the ring centre of symmetry from the star allows us to quantify its eccentricity to be $e=0.09_{-0.01}^{+0.02}$. This eccentric feature is also detected in low resolution \textit{Herschel}/PACS observations, under the form of a pericenter-glow. Combining the infrared and millimeter photometry, we retrieve a disk grain size distribution index of $\sim -3.4$, and therefore exclude in-situ formation of the inferred belt-shaping perturber, for which we provide new dynamical constraints. Finally, ALMA images show four point-like sources that exceed 100$\,\mu$Jy, one of them being just interior to the ring. Although the presence of a background object cannot be excluded, we cannot exclude either that this source is circumplanetary material surrounding the belt-shaper, in which case degeneracies between its mass and orbital parameters could be lifted, allowing us to fully characterize such a distant planet in this mass and age regime for the very first time.

\end{abstract}


\section{Introduction} \label{sec:intro}

Debris disks contain solid bodies in a collisional cascade, ranging from km-sized down to micron-sized dust grains. They are remnants of planetary formation processes \citep[see, e.g., the review by][]{Krivov2010}. As examples, our own Solar System hosts the Main Asteroid and the Kuiper belts. Extrasolar debris disks were initially detected by the \textit{InfraRed Astronomical Satellite} (IRAS) through the infrared excess that the micron-sized dust grains add to their host star's emission \citep{Aumann1984}.
Since the top reservoir of km-sized bodies is not expected to be replenished, it is expected that debris disks lose luminosity with time, until they become undetectable with our current instruments, and as confirmed by observations \citep{Wyatt2008,Sierchio2014}. Consequently, debris disks are rarely detected and even more rarely resolved in systems with ages comparable to that of the Solar System. Opportunities to investigate the outcome and diversity of mature planetary systems are therefore scarce. One such opportunity is the system of HD 202628.

HD 202628 (HIP 105184) is a G2V star, located at 23.8 pc \citep{Collaboration2016,Collaboration2018,Bailer-Jones2018}, and has an estimated age of $1.1\pm0.4$ Gyr (see Appendix \ref{app:eric}). 
The debris disk of HD 202628 was first revealed with \textit{Spitzer}, which detected a significant 70 microns excess of nearly 20 times the star's photospheric flux at this wavelength, and with fractional infrared luminosity $1.4 \times 10^{-4}$ \citep{Koerner2010}. It later appeared extremely well resolved in visible scattered light with the \textit{Hubble Space Telescope}'s Space Telescope Imaging Spectrograph (\textit{HST}/STIS), and showed a cleared central zone approximately 6$\arcsec$ in radius, a sharp inner edge, while extending outwards to at least 9$\arcsec$ in radius. Most importantly, the ring was found to be eccentric: it was found to extend further from the star on its South-East ansa than to the North West one, that is, the star was found offset from the projected ring center \citep{Krist2012,Schneider2016}. 

Extrasolar debris disks have been found to bear imprints of interactions with planets, as often revealed by asymmetries in their spatial distribution \citep[][]{Wyatt1999a,Krivov2010}. With its sharp inner edge and eccentric shape, the morphology of the debris disk of HD 202628 is analogous to that hosted by the mature yet younger \citep[$\sim 440\,$Myr, ][]{Mamajek2012a} A-type star Fomalhaut \citep{Kalas2005}, around which the presence of a planet on an eccentric orbit at several tens of AU, and carving the inner edge of the ring has been subsequently inferred through dynamical modeling \citep{Quillen2006,Chiang2009}. Consequently, a similar perturber has been postulated around the Gyr-old solar-type star HD 202628 \citep{Krist2012}.

Dynamical modelling of debris disk asymmetries and structures allows us to predict the presence of yet undetected planetary components \citep{Moro-Martin2013}. This approach permits us to constrain their mass and orbital properties, and hence constitutes a powerful indirect detection technique, which in addition, probes regions of exoplanet mass and orbital radius parameter space inaccessible to other methods. Indeed, it is particularly suited for systems such as Fomalhaut or HD 202628, where planets are too distant from their host star to have been detected through usual detection techniques, such as radial velocities or transits, which are biased towards short-period objects. 

\begin{figure}
\centering
\makebox[\columnwidth]{\includegraphics[scale=0.6]{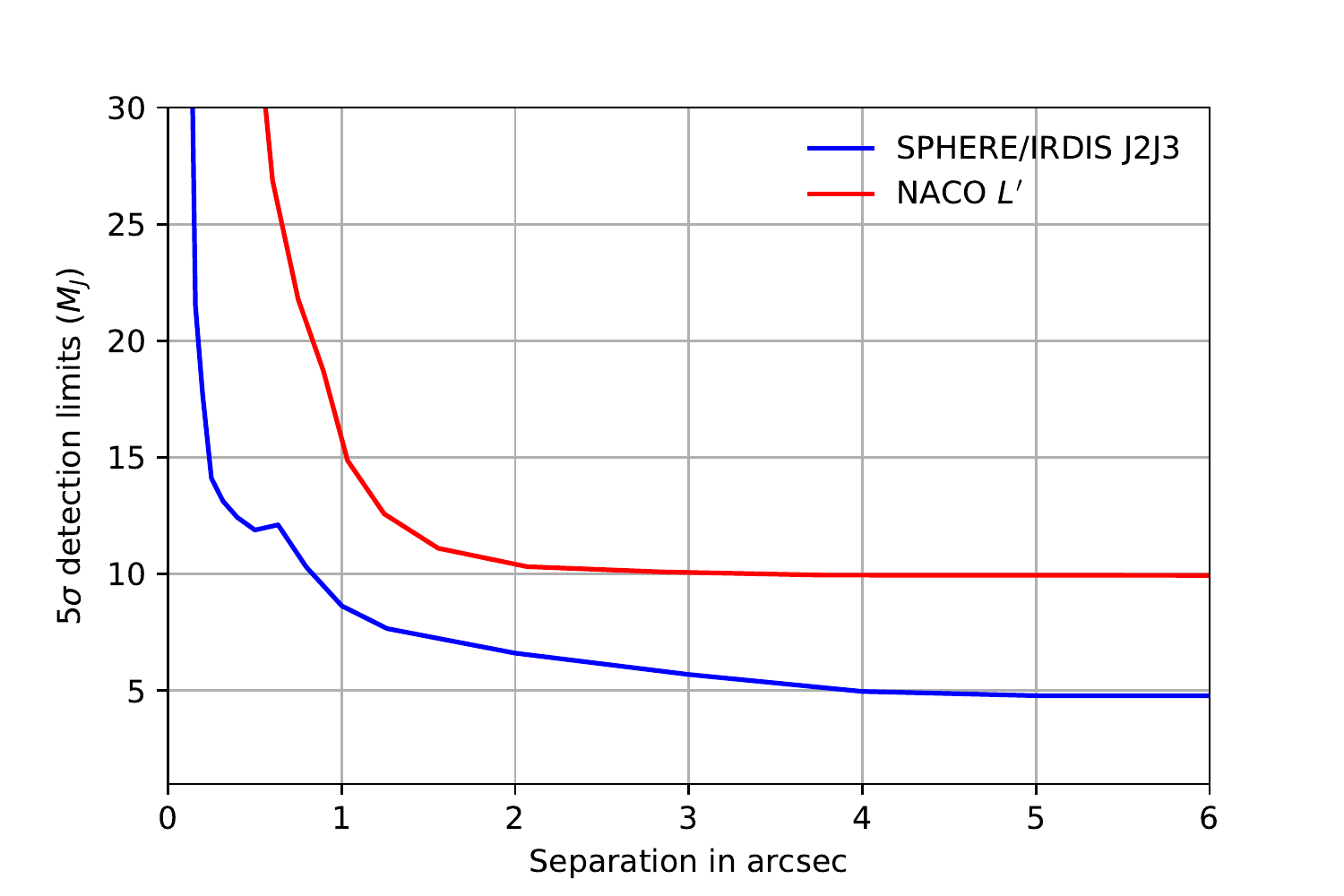}}
\caption{Five sigmas detection limits in mass reached by Irdis and NaCo around HD 202628, as a function of the separation to the host star, assuming an age of 1.1 Gyr and using the COND models of \citet{Baraffe2003}.}
\label{fig:detection_limits}
\end{figure}

While direct imaging techniques and instruments may overcome these limitations, in mature systems such as HD 202628, planets are expected to have lost too much of their intrinsic luminosity to be detected. Around a star as old as HD 202628, and even with the best configuration possible to probe very cold companions \citep[for example, see the Very Large Telescope's Spectro-Polarimetric High-Contrast Exoplanet Research (VLT/SPHERE) observations of GJ 504A in dual band imaging J2J3;][]{Bonnefoy2018}, the high contrast between such companions and HD 202628 would make it difficult to directly observe objects less massive than 5\,$\mathrm{M_{Jup}}$ at the inferred distance (see Figure~\ref{fig:detection_limits}). Note that current direct imaging constraints for this system obtained with VLT/NaCo (NAOS-CONICA; Nasmyth Adaptive Optics System (NAOS) Near-Infrared Imager and Spectrograph (CONICA)) in L band allow us to exclude the presence of a body more massive than $\sim 50\mathrm{M_{Jup}}$ beyond $\sim 1\arcsec.5$ (Mawet, private communication).

Based on the geometrical constraints provided by \textit{HST} for the debris disk of HD 202628, several predictions have been made on the putative eccentric belt-shaping perturber at play in this system. Applying their equation that links the width of a debris ring as seen in scattered light to the maximum mass and minimum semimajor axis of a belt-shaping perturber, \citet{Rodigas2014} theoretically derived a perturber's maximum mass of 15\,$\mathrm{M_{Jup}}$ at minimum separation $\sim1\arcsec.2$.
On the other hand, based on dynamical theoretical constraints for an eccentric perturber to secularly shape a ring into an eccentric ring and carving its inner edge, \citet{Pearce2014} provided a lower mass limit of 0.2\,$\mathrm{M_{Jup}}$, while further extensive numerical exploration and fit to \textit{HST} observations allowed \citet{Thilliez2016} to find a best fit of 3\,$\mathrm{M_{Jup}}$ for this eccentric companion, which means this perturber should have a mass well below the current achievable detection limits quoted above. This makes it tough to characterize it other than by carrying out dynamical modelling of the spatial structure of the ring, and thus characterizing it via its gravitational imprint on the debris disk. The aforementioned constraints were derived using the star's distance as provided by \textit{Hipparcos} \citep[d=24.4pc,][]{Leeuwen2007} which was recently corrected by \textit{Gaia} measurements, though the difference between the two values is relatively small and will therefore not affect the constraints significantly. However, the constraints were also based on the geometry of the debris disk as found by \textit{HST} observations: small micron-sized particles revealed in scattered light with \textit{HST} can drift far from their source due to the influence of collisions and radiation pressure \citep{Thebault2007}, which leads structures seen in scattered light to be altered versions of the true dynamical structure imposed by an unseen planet \citep{Thebault2012}.

Knowledge of the debris disk geometry should be clearly improved thanks to far-IR and up to millimeter wavelengths observations, as those probe a different physical regime than visible light: the large mm-sized particles traced by these emissions tend to remain near the site of their initial collisional formation. This means that millimeter observations of a debris disk provide better constraints on the gravitational imprint of the planetary components shaping them. In addition, comparison of a debris disk morphology and content across different wavelengths provides key information on the dust population, and in particular, its size distribution \citep{Ricci2012}. This is of prime importance in the case of the HD 202628 system as in-situ formation of the belt-shaping perturber has been suggested to be possible, and promoted by a steeper grain size distribution than usually expected in debris disks \citep{Kenyon2015}. 

We present here a whole new set of observations of the debris disk of HD 202628, which we will describe in Section \ref{sec:obs}, and analyze in Section \ref{sec:results}: Atacama Large Millimeter/submillimeter Array (ALMA) resolved observations at 1.3 mm which will allow us to significantly refine our knowledge of the parent ring geometry while probing the disk gas content, \textit{Herschel}/PACS (Photodetector Array Camera and Spectrometer) and SPIRE (Spectral and Photometric Imaging Receiver) observations from 70 to 500 microns, which will provide photometric measurements and allow us, in combination with our ALMA data, to determine the grain size distribution of the debris disk and conclude on the possible in-situ formation of distant planets around HD 202628, and finally, deeper \textit{HST}/STIS observations in scattered light, which will provide additional information on the dust grain properties and in particular, their scattering phase function. Using the new constraints on the disk geometry provided by our observations, we will set in turn constraints on the orbital properties of the eccentric perturber inferred in this system in Section \ref{sec:planet}. In addition, we will discuss the presence of a source just interior to the ring which, if linked to the system, could be circumplanetary material orbiting the expected perturber, and show how this would allow us to fully characterize the mass and orbital properties of the perturber. Finally, we present our conclusions in Section \ref{sec:conclu}.  


\section{Observations}\label{sec:obs}

\subsection{ALMA Band 6 Observations}\label{sec:ALMA_obs}

We present here ALMA observations of HD 202628 in Band 6  (230 GHz, 1.3mm), under the project 2016.1.00515.S (PI: V. Faramaz). They comprise 12m-array observations carried out from 2017 April 15 to 2017 April 29, and observations carried out with the Atacama Compact Array (ACA) from 2016 October 19 to 2016 November 2.
Our ACA and 12m-array data comprised 13 and 7 separate observations, respectively, for which we summarize the characteristics in Table \ref{tab:obslogACA} and \ref{tab:obslog12m}. 
ACA data were taken using baselines ranging from 8.9 to 48.0 m, which corresponds to angular scales of $30\arcsec.2$ and $5\arcsec.6$, respectively, while 12m-array data were taken using baselines ranging from 15.1 to 460.0 m, which in this case, corresponds to angular scales of $17\arcsec.8$ and $0\arcsec.6$, respectively. Given the distance of the star (23.8 pc), this means that the spatial scales that were probed with the ACA ranged from 133.3 to 718.8 AU, and those probed with the 12m-array ranged from 14.3 to 423.6 AU. 
The spectral setup consisted of four spectral windows, each 2 GHz-wide. Three were centred on 232.5, 245.5, and 247.5 GHz, and divided into 128 channels of width 15.625 MHz $(\sim 20\,\mathrm{km.s^{-1}})$. 

Although we did not expect primordial gas to be present in a system as old as HD 202628, we nevertheless used the fourth spectral window to probe CO gas, via the J=2-1 emission line, as these can be released from collisions among planetesimals \citep[as for instance, in the mature systems of $\eta$ Corvi and Fomalhaut;][, respectively]{2017MNRAS.465.2595M,Matra2017a}. Therefore, the fourth spectral window was centred on 230.5 GHz, with a large number of finer channels (4096 for the 12m array observations and 3840 for the ACA observations), leading to a spectral resolution of 0.5 MHz $(\sim 0.6\,\mathrm{km.s^{-1}})$. The total time on source was 9.8 hours with the ACA and 5.3 hours with the 12m array.

The spectral window covering the CO J=2-1 transition (230.538 GHz, with a spectral channel width of 488.24 kHz, or $0.63\,\mathrm{km.s^{-1}}$) was extracted from each of the 12m array observations, and combined to produce a visibility dataset for CO imaging. We subtracted continuum emission from the visibilities using the \textit{uvcontsub} task in CASA v5.1.0. Then, we produced a dirty image of the CO dataset using the \textit{tclean} CASA task, with natural weighting. We find no clear emission at or near the radial velocity of the star in the data cube, which has an RMS sensitivity of 0.36 mJy.beam$^{-1}$ in a $0.63\,\mathrm{km.s^{-1}}$ channel, for a synthesized beam size of $0.96\arcsec \times 0.77\arcsec$. 


\begin{deluxetable*}{ccccccccc}[htbp] 
\tablecaption{Summary of our ALMA/ACA observations at 1.3 mm (Band 6). \label{tab:obslogACA}}
\tablewidth{0pt}
\tablehead{
\colhead{Date\tablenotemark{a}} &
\colhead{Time\tablenotemark{a}} &
\colhead{On source} & \colhead{$\mathrm{N_{Ant.}}$} & \colhead{PWV} & \colhead{Elevation} & \multicolumn{3}{c}{Calibrators} \\
\colhead{(YYYY-mm-dd)} & \colhead{(UTC)} &
\colhead{(min)} & \colhead{} & \colhead{(mm)} & \colhead{(deg)} & \colhead{Flux} & \colhead{Bandpass} & \colhead{Phase}
}
\startdata
2016-10-19 & 01:38:51.0 & 12.8 & 10 & 0.68-0.73 & 54.9-57.0 & Uranus & J0006-0623 & J2056-4714  \\[2pt]
2016-10-21 & 01:23:11.2  & 48.1 & 9  & 0.39-0.48  & 47.5-58.2 & Neptune & J0006-0623 & J2056-4714  \\[2pt]
2016-10-24 & 01:17:29.5  & 48.1 & 8 & 0.62-1.16  & 46.2-57.2 & Neptune & J0006-0623 & J2056-4714  \\[2pt]
2016-10-27 & 00:38:56.5  & 48.1 & 10 & 0.36-0.55  & 51.0-61.3 & Neptune & J0006-0623 & J2056-4714  \\[2pt]
2016-10-27 & 23:38:25.1& 48.1 & 9 & 0.68-1.07 & 60.5-68.0 & Mars & J1924-2914 & J2056-4714  \\[2pt]
2016-10-28 & 01:17:52.8 & 48.1 & 9 & 0.63-0.70 & 43.5-54.9 & Neptune & J0006-0623 & J2056-4714  \\[2pt]
2016-10-28 & 23:36:45.8 & 48.1 & 10 & 0.76-1.03 & 60.1-67.9 & Mars & J1924-2914 & J2056-4714  \\[2pt]
2016-10-29 & 23:48:46.8 & 48.1 & 10 & 1.45-1.85 & 57.5-66.4 & Mars & J1924-2914 & J2056-4714  \\[2pt]
2016-10-30 & 01:30:46.1 & 48.1 & 10  & 1.08-1.26 & 39.9-51.0 & Uranus & J0006-0623 & J2056-4714  \\[2pt]
2016-10-30 & 23:30:23.8 & 48.1 & 10  & 0.65-0.80 & 59.8-67.7 & Mars & J1924-2914 & J2056-4714  \\[2pt]
2016-10-31 & 01:21:14.7 & 48.1 & 10  & 0.44-0.63 & 40.6-52.0 & Uranus & J0006-0623 & J2056-4714  \\[2pt]
2016-11-01 & 00:02:10.6 & 48.1 & 10  & 0.88-0.94 & 54.0-63.8 & Mars & J1924-2914 & J2056-4714  \\[2pt]
2016-11-01 & 22:56:59.3 & 48.1 & 10 & 1.46-1.66 & 63.5-69.4 & Mars & J1924-2914 & J2056-4714  \\[2pt]
\enddata
\tablenotetext{a}{At exposure start.}
\end{deluxetable*}

\begin{deluxetable*}{ccccccccc}[htbp] 
\tablecaption{Summary of our ALMA/12m-array observations at 1.3 mm (Band 6). \label{tab:obslog12m}}
\tablewidth{0pt}
\tablehead{
\colhead{Date\tablenotemark{a}} &
\colhead{Time\tablenotemark{a}} &
\colhead{On source} & \colhead{$\mathrm{N_{Ant.}}$} & \colhead{PWV} & \colhead{Elevation} & \multicolumn{3}{c}{Calibrators} \\
\colhead{(YYYY-mm-dd)} & \colhead{(UTC)} &
\colhead{(min)} & \colhead{} & \colhead{(mm)} & \colhead{(deg)} & \colhead{Flux} & \colhead{Bandpass} & \colhead{Phase}
}
\startdata
2017-04-15 & 11:34:48.2 & 46.2 & 42 & 1.86-2.34 & 68.6-69.6 & Titan & J2056-4714 & J2056-4714  \\[2pt]
2017-04-18 & 11:24:08.9 & 46.2 & 41 & 1.47-1.56  & 68.2-69.6 & Titan & J2056-4714 & J2056-4714  \\[2pt]
2017-04-24 & 10:04:34.5  & 38.9 & 39 & 0.71-0.74  & 64.0-68.0 & Titan & J2258-2758 & J2056-4714  \\[2pt]
2017-04-25 & 11:21:26.5  & 46.2 & 38 & 1.10-1.44  & 66.5-69.5 & J2056-4714 & J2258-2758 & J2056-4714  \\[2pt]
2017-04-27 & 10:56:34.8 & 46.2 & 39  & 0.27-0.28 & 67.6-69.6 & Titan & J2056-4714 & J2056-4714  \\[2pt]
2017-04-28 & 08:29:20.3 & 46.2 & 39 & 0.24-0.26 & 51.0-60.0 & J2056-4714 & J1924-2914 &J2056-4714 \\[2pt]
2017-04-29 & 11:08:34.5 & 46.2 & 39 & 0.78-0.93 & 66.0-69.5 & J2056-4714 & J2258-2758 & J2056-4714  \\[2pt]
\enddata
\tablenotetext{a}{At exposure start.}
\end{deluxetable*}


\subsection{\textit{Herschel}/PACS and SPIRE Observations}\label{sec:hst_obs}

\textit{Herschel}/PACS observations using the mini scan map AOT (Astronomical Observing Template) took place on 2012 March 28. Ten scan legs with length $3\arcmin$ were executed 3 times, repeated in two concatenated AORs (Astronomical Observation Request) scanning along the PACS array diagonal of 70 and 110 degrees.  This sequence was done once for simultaneous 70 and 160 $\mu$m observations, and a second time for simultaneous 100 and 160 $\mu$m observations.  The scans were filtered to remove pattern noise, excluding the region around the target, and rendered into mosaics with $1\arcsec.0$, $1\arcsec.0$, and $2\arcsec.0$ pixels (at 70, 100, 160 $\mu$m) using customized routines in Version 10 of the \textit{Herschel} Interactive Processing Environment (HIPE) software package. 
\textit{Herschel}/SPIRE  observations took place on 2012 May 11 using five repetitions of the small map AOT in each of the three photometric bands. Processed mosaics with  $6\arcsec$, $10\arcsec$, and $14\arcsec$ pixels (at 250, 350, 500 $\mu$m) were retrieved from the \textit{Herschel} Archive.


\subsection{New \textit{HST}/STIS observations}\label{sec:herschel_obs}

HD 202628 has been observed with the \textit{HST}/STIS coronagraph $(0."05\,\mathrm{pixel}^{-1})$ in  three separate Guest Observer (GO) programs (Table \ref{tab:hst_obslog}). The first, GO-12291 (PI=Krist), was a 20-orbit imaging survey of 10 stars with infrared excesses measured by the  \textit{Spitzer Space Telescope}. Each target was observed over two consecutive orbits, with the telescope rolled about the star by 28$^{\circ}$. Of those, a disk was seen only around HD 202628 \citep{Krist2012}. The extreme faintness of the disk in the 2011 data meant that additional integrations were needed  to better define its morphology. A follow-up program, GO-13455 (PI=Krist), obtained 9 more orbits of images at 9 different orientations spread over three epochs in 2014, which are described here for the first time. Finally, a separate program that revisited previously-imaged debris disks (GO-13786, PI=Schneider) obtained 6 more orbits of data at 6 orientations at two epochs in 2015 \citep{Schneider2016}.

Due to the large brightness difference between a star and its debris disk, the surface brightness of the wings of the star's instrumental point spread function (PSF) is typically greater than the disk's. A coronagraph is used to suppress the diffraction pattern caused by the telescope's obscurations; a deformable mirror, if present, can be used to further reduce the starlight by correcting for optical aberrations that scatter stellar flux into the PSF wings (though there is not one on any \textit{HST} instrument). The residual starlight is then removed from the images using some sort of post-processing technique. The simplest, reference differential imaging (RDI), is the subtraction of a reference star image from the target star's. It has been used extensively on \textit{HST} and ground-based high-contrast data. Another technique is angular differential imaging (ADI), which extracts the astronomical signal (\textit{e.g.}, disk) by  observing the field at  multiple orientations and solving for what moves (the sky) and what does not (the PSF).

The STIS coronagraph uses a crossed pair of wedges to block the star at an intermediate focal plane, with the wedge position chosen to allow imaging as close to the star as desired (the wider wedge positions offer better diffraction suppression). At a subsequent pupil in the optical train, a Lyot stop mask is used to suppress diffraction from the outer edge of the telescope aperture. Because the STIS Lyot stop does not mask the telescope secondary mirror or its support spiders, diffraction spikes remain visible in the coronagraphic images and the wings of the PSF are suppressed by only a factor of a few \citep{Krist2004}. Lacking a deformable mirror, the remaining starlight must be removed using post-processing. PSF subtraction provides the greatest gain in starlight suppression with \textit{HST} due to its relatively stable (compared to the ground) PSF, rather than the suppression provided by the unoptimized coronagraph alone.

The STIS coronagraph does not have any filters, so its bandpass is effectively limited by the wavelength response of the CCD detector, which spans over 250 - 1100 nm. This complicates finding suitable reference PSF stars for RDI post-processing. The stellar diffraction pattern varies with wavelength, so small differences in the spectral energy distributions (SEDs) between the target and reference stars can lead to mismatches in the PSFs and cause residual artifacts in the processed images. The broader the bandpass, the greater the effects from SED mismatches will be, so a STIS reference star must be chosen that matches the target's star color extremely well.

In the original GO-12291 survey by Krist et al., the desire to avoid the (at a minimum) 10 additional orbits needed for the 10 matching reference stars led to the use of ADI post-processing with multi-orientation observations instead of RDI. Besides saving orbits, ADI avoids the introduction of residuals due to stellar color differences. This same technique was used to first image the disk around HD 207129 \citep{Krist2010}; the same iterative ADI algorithm used here was described in that paper. The drawback of the ADI method for an extended object like a disk is that self-subtraction will likely occur, depending on the object's morphology. For example, a face-on disk would appear to be part of the PSF since it does not appear to move with telescope orientation, so it would not appear in the processed image. For less inclined, ring-shaped disks like HD 207129 and HD 202628, the self-subtraction can be reduced by maximizing the amount of orientation change and the number of orientations.

Prior to post-processing, all of the images were retrieved from the \textit{HST} archive. Although the calibration pipeline combines, with cosmic-ray rejection, the flat-fielded subexposures  $(\mathrm{\_flt.fits})$ into final images  $(\mathrm{\_crj.fits})$, no image registration is done. For the best post-processing results, however, this is needed,  especially since the PSF structure has high-spatial-frequency streaks. By subtracting one frame from another the residuals  caused by drift of the star within an orbit can be seen as an oversubtraction on one side of the star and undersubtraction  on the other, especially in the diffraction spikes. Using cubic convolution interpolation, each subexposure was iteratively  shifted by subpixel amounts and subtracted from the first 2011 subexposure until the residuals appeared to be visually minimized.  Shifts as low as 0.02 pixels (1 mas) produced noticeable results. The aligned subexposures for each orientation were then combined with cosmic ray rejection.

The image sets from 2011, 2014, and 2015 were separately ADI processed (for consistency, the 2011 data were reprocessed). This avoided the long-term changes in the PSF. Also, the 2015 data were taken at a different wedge position than the 2011 and 2014 sets, and the PSF is different for each position. Note that the ADI processing of the 2015 data and the derived results presented here are independent of those for the same data discussed by Schneider et al. (2016), who used RDI instead. The center of rotation in the aligned images was determined by fitting for the intersection of the diffraction spikes. The combination of these three images are used hereafter, representing a total of 8.9 hours of integration. An evaluation of the impact of disk self-subtraction due to ADI processing using a disk model is given in the Appendix \ref{app:john}. As will be shown, there is no significant effect that would alter the observed morphology of the disk.

\begin{deluxetable*}{cccccccc}[htbp] 
\tablecaption{List of \textit{HST} HD 202628 Observations. \label{tab:hst_obslog}}
\tablewidth{0pt}
\tablehead{
   &   & \colhead{Total}  &  &  &  &  &  \\[5pt]
\colhead{Date}  & \colhead{\textit{HST}} & \colhead{Exposure}  & \colhead{Number of} & \colhead{CCD} &  &  &\colhead{Program} \\[5pt]
  YYYY-MM-DD & \colhead{Dataset}        & \colhead{Time}         & \colhead{Subexposures}  & \colhead{Gain}  & \colhead{Occulter} & \colhead{Orientation\tablenotemark{a}} & \colhead{Number}
}
\startdata
2011-05-15 & OBHS05010 & 2256 s & 8 & 1 & WEDGEA1.8 & -165.0$^{\circ}$ & 12291 \\[2pt]
2011-05-15 & OBHS06010 & 2256 s & 8 & 1 & WEDGEA1.8 & -137.0$^{\circ}$ & 12291 \\[2pt]
2014-05-02 & OC8F01010 & 1976 s & 8 & 1 & WEDGEA1.8 & -160.5$^{\circ}$ & 13455 \\[2pt]
2014-05-02 & OC8F02010 & 1976 s & 8 & 1 & WEDGEA1.8 & -145.5$^{\circ}$ & 13455 \\[2pt]
2014-05-02 & OC8F03010 & 1976 s & 8 & 1 & WEDGEA1.8 & -130.5$^{\circ}$ & 13455 \\[2pt]
2014-07-12 & OC8F04010 & 1976 s & 8 & 1 & WEDGEA1.8 & -115.5$^{\circ}$ & 13455 \\[2pt]
2014-07-12 &OC8F05010 & 1976 s & 8 & 1 & WEDGEA1.8 & -100.5$^{\circ}$ & 13455 \\[2pt]
2014-07-12 & OC8F06010 & 1976 s & 8 & 1 & WEDGEA1.8 & -85.5$^{\circ}$ & 13455 \\[2pt]
2014-08-07 & OC8F07010 & 1976 s & 8 & 1 & WEDGEA1.8 & -70.5$^{\circ}$ & 13455 \\[2pt]
2014-08-07 & OC8F08010 & 1976 s & 8 & 1 & WEDGEA1.8 & -55.5$^{\circ}$ & 13455 \\[2pt]
2014-08-07 & OC8F09010 & 1976 s & 8 & 1 & WEDGEA1.8 & -40.5$^{\circ}$ & 13455 \\[2pt]
2015-09-17 & OCJC01040 & 1630 s & 5 & 4 & WEDGEA1.0 & -18.0$^{\circ}$ & 13786 \\[2pt]
2015-09-17 & OCJC02040 & 1630 s & 5 & 4 & WEDGEA1.0 & 1.6$^{\circ}$ & 13786 \\[2pt] 
2015-09-17 & OCJC04040 & 1630 s & 5 & 4 & WEDGEA1.0 & 21.0$^{\circ}$ & 13786 \\[2pt]
2015-05-30 & OCJC05040 & 1630 s & 5 & 4 & WEDGEA1.0 & -117.0$^{\circ}$ & 13786 \\[2pt]
2015-05-30 & OCJC06040 & 1630 s & 5 & 4 & WEDGEA1.0 & -140.0$^{\circ}$ & 13786 \\[2pt]
2015-05-30 & OCJC08040 & 1630 s & 5 & 4 & WEDGEA1.0 & -163.0$^{\circ}$ & 13786 \\[2pt]
\enddata
\tablenotetext{a}{Angle from North through East to the +Y image axis as reported by the ORIENTAT image file header keyword.}
\end{deluxetable*}


\section{Results}\label{sec:results}

\subsection{ALMA Band 6 Observations}\label{sec:ALMA_res}

\begin{figure*}
\makebox[\textwidth]{\includegraphics[scale=0.5]{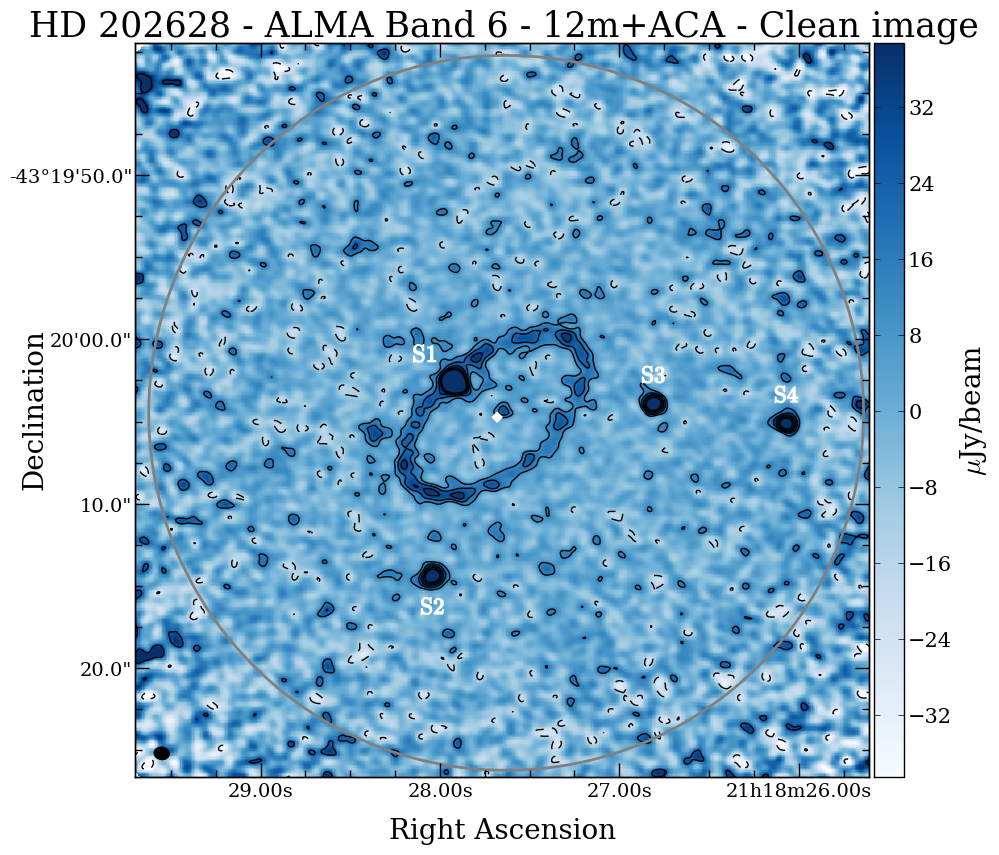}}
\caption{ALMA 1.3 continuum observations of HD 202628, combining the 12m and ACA data. Contours show the $\pm$2, 4, 6,... $\sigma$ significance levels, with $\sigma=5.2\,\mu\mathrm{Jy.beam}^{-1}$. The synthesized beam, shown on the lower left side of the image, has dimensions $0\arcsec.92 \times 0\arcsec.75$, with position angle $83^{\circ}$. The stellar photosphere appears inner to the ring and is detected with SNR of nearly 6, and the ring center of symmetry (white diamond) is slightly offset from it. Bright sources (SNR$>6$) within the field are labeled from S1 to S4. The grey circle indicates the 50\% response of the ACA primary beam and the color bar shows the fluxes in $\mu\mathrm{Jy.beam}^{-1}$.}
\label{fig:HD202628_continuum}
\end{figure*}

Calibrations were applied using the pipeline provided by ALMA. We used the \emph{TCLEAN} algorithm in CASA (Common Astronomy Software Applications) version 5.1.1 \citep{McMullin2007} to perform the image reconstruction of the continuum emission, that is, to obtain the inverse Fourier transform of the observed visibilities. We combined the four spectral windows in order to recover the maximum signal-to-noise ratio (SNR) and used a mask generated with our best fit model (see Section ~\ref{sec:ALMA_ring}). 
We used a natural weighting scheme, a cell size of $0\arcsec.08$, and an image size of $800\times 800$ pixels in order to cover the primary beam, as well as a mask generated with our disk best fit model (see Section \ref{sec:ALMA_ring}). The resulting synthesized beam has dimensions $0\arcsec.92 \times 0\arcsec.75$, with position angle (PA) $83^{\circ}$. The rms was measured in a large region far from the sources present in the field of view, and was found to be $\sigma=5.2\,\mu\mathrm{Jy.beam}^{-1}$. We further corrected the image for the primary beam, and show the results in Figure \ref{fig:HD202628_continuum}. 

The ring appears clearly defined in our image, as well as the star, that is visibly offset from the ring centre of symmetry. In addition, several bright sources appear within the field, one of them being superimposed on the ring. In the following, we give quantitative measurements for all these components: ring's photometry and geometry, stellar photospheric flux and position, bright sources' photometry and position, and finally gas content.

\subsubsection{The star}\label{sec:ALMA_star}

The peak emission of the unresolved stellar photosphere at 1.3mm is found to be $29.3\pm 5.2\mu\mathrm{Jy.beam}^{-1}$.
This is higher than our prediction of a star's peak flux emission of $19\mu\mathrm{Jy.beam}^{-1}$, which was derived by fitting a Kurucz model photosphere to the stellar photometry between 1-24 microns (see Section \ref{sec:SED}). With $\sigma=5.2\mu\mathrm{Jy.beam}^{-1}$, HD 202628 is detected with a SNR of nearly 6, which allows us to constrain its position. A Gaussian ellipse fitted to the photosphere's emission is centred on $21^{\mathrm{h}}18^{\mathrm{m}}27^{\mathrm{s}}.653 \pm 0.004$ in Right Ascension, and $-43^{\circ}20\arcmin 04\arcsec .267 \pm 0.04$ in Declination\footnote{Note that the astrometric accuracy returned by the Gaussian fitting procedure is a factor of 2-3 smaller than ALMA's nominal astrometric accuracy given by $70000\times(\nu \times B \times SNR)^{-1}$, where $\nu$ is the frequency of the observations (in GHz) and $B$ is the largest used baseline (in km) (Equation (10.7) of Cycle 7 ALMA Technical Handbook).}.

Being able to detect the star's emission and constrain its position is crucial in order to determine the geometry of the debris ring. Since the debris ring has been found to be eccentric in \textit{HST} data, its center of symmetry was expected to be offset from the star, as is already visible in our image. Knowledge of the star position allows us to measure this offset and retrieve constraints on the ring eccentricity (see Section \ref{sec:ALMA_ring}). 

\newpage

\subsubsection{Bright sources}\label{sec:BGsources}

Apart from the disk and the star, the field contains four additional bright sources, for which we retrieve the extent, brightness, and position. For each source, and within the CASA viewer, we performed an elliptical Gaussian fit over a rectangular region encompassing the source. This procedure returns the best fit's centre, its peak and integrated fluxes, along with its convolved elliptical dimensions and position angle. If the source is found resolved, unconvolved ellipse characteristics are returned as well by the fit. We summarize this information in Table \ref{tab:sources}. The brightest source (labeled S1 in Figure \ref{fig:HD202628_continuum}), appears within the ring, North-East to the star, while the other three sources (labeled S2, S3, and S4) are exterior to the ring. Note that we used an image from which our disk best fit model (see Section \ref{sec:ALMA_ring}) was subtracted, so as to disentangle the emission of the source S1 from that of the ring. All the sources were marginally resolved and, with $\sigma=5.2\,\mu\mathrm{Jy/beam}^{-1}$, were found to have SNRs 36, 17, 19, and 22, fro S1, S2, S3, and S4 respectively.

\begin{deluxetable*}{cccccccc}[htbp] 
\tablecaption{Characteristics of the bright sources within the field of view. These measurements have been made using an elliptical Gaussian fit to the sources. \label{tab:sources}} 
\tablewidth{0pt}
\tablehead{
\colhead{Source}  & \colhead{Integrated} & \colhead{Peak}  & \multicolumn{2}{c}{Position\tablenotemark{a}} & \multicolumn{3}{c}{Deconvolved FWHM axes} \\[5pt]
         & \colhead{Flux $(\mu\mathrm{Jy})$} & \colhead{$(\mu\mathrm{Jy/beam}^{-1})$} & \colhead{Right Ascension (s)}  & \colhead{Declination ($\arcsec$)}  & \colhead{Major (mas)} & \colhead{Minor (mas)} & \colhead{PA $(^\circ)$}
}
\startdata
S1 & $285\pm18$ & $189.0\pm 7.6$    & $21:18:27.905 \pm 0.002$  & $-43.20.02.629 \pm 0.017$ & $700 \pm 108$ & $460 \pm 153$ & $28 \pm 22$ \\[2pt]
S2 & $154\pm12$  & $89.9\pm 4.6$  & $21:18:28.030 \pm 0.003$ & $-43.20.14.379 \pm 0.020$  & $851\pm 124$  & $ 553\pm 138$ & $121\pm 21$  \\[2pt]
S3 & $123\pm10$  & $100.0\pm 5.0$ & $21:18:26.798 \pm 0.002$  & $-43.20.03.896 \pm 0.018$ & $456\pm 139$  & $299\pm 294$  & $8\pm 77$  \\[2pt]
S4 & $150\pm11$  & $114.7\pm5.4$ & $21:18:26.059 \pm 0.002$ & $-43.20.05.077 \pm 0.017$  &$514 \pm 139$  & $405 \pm 194$ & $52\pm63$  \\[2pt]
\enddata
\tablenotetext{a}{As compared with ALMA's nominal astrometric accuracy (given by Equation (10.7) of Cycle 7 ALMA Technical Handbook as $70000\times(\nu \times B \times SNR)^{-1}$, where $\nu$ is the frequency of the observations (in GHz) and $B$ is the largest used baseline (in km)), the astrometric accuracy returned by the Gaussian fitting procedure is consistent for the sources observed at the largest SNR (that is, S1 and S4), while it is smaller by a factor of two for the sources S2 and S3.}
\end{deluxetable*}

\subsubsection{The ring}\label{sec:ALMA_ring}

As compared to the \textit{HST} images of \citet{Krist2012} and \citet{Schneider2016}, the debris disk appears as a narrow ring ($\sim 1"$ wide) with ALMA. It is contained within two projected ellipses of dimensions $\sim 3" \times \sim 6"$ and $\sim 4" \times \sim 7"$, both with position angle of $\sim -50^\circ$.

More importantly, their center of symmetry is offset by $\sim 0".5$ from the star, which confirms the ring is intrisically eccentric.

In order to further characterize the ring's geometry, orientation, and mass, we used the Python \texttt{emcee} procedure \citep{Foreman-Mackey2013}, which allows us to perform an MCMC (Markov Chain Monte Carlo) fit of a debris disk model emission to our observations. The model disk emission is computed thanks to the ray-tracing disk code first developed by \citet{Rosenfeld2013} and \citet{Flaherty2015} to model protoplanetary disks emission, and later modified to model debris disks emission, which are optically thin in comparison with optically thick protoplanetary disks \citep{Daley2019}\footnote{This code is publicly available and can be found here: \url{https://github.com/kevin-flaherty/disk_model}}. 

The dust temperature is derived assuming blackbody grains irradiated by the star, while its opacity is set to $2.3\mathrm{cm^2/g}$ \citep{Beckwith1990,Andrews2005,Andrews2013}. This opacity has been originally derived from an empirical power-law inspired by dust grain distribution models, and hence there is an assumption on the dust composition and grain size distribution underlying it. However, it shall be noted that there remains a factor of five uncertainty on the opacity normalization, and more importantly, that the dust grain size distribution does not extend up to the parent planetesimal population of the debris disk. Consequently, the mass derived by the radiative transfer code corresponds to the mass of the grains seen at the wavelength of the observations ($\sim 1\,$mm), and the parent bodies could have much more mass.

We used a circular ring model, which treated the offset of its center of symmetry from the centre of our image with two free parameters (Right Ascension and Declination).
From the original \textit{HST} data, the inferred eccentricity of the ring is close to 0.2 \citep{Krist2012}. Since the semiminor axis $b$ of an ellipse is by definition $b=a\sqrt{1-e^2}$, where $a$ is its semimajor axis and $e$ its eccentricity, then the difference between the major and minor axis is expected to be $\Delta=a-b=a(1-\sqrt{1-e^2})$, that is, $\sim 2\%$ of the semimajor axis. Using the largest extent of the ring of $\sim 250\,$AU observed with \textit{HST} as a conservative value, then the minor axis is expected to be $\lesssim 0".2$ smaller than the major axis. This is four times smaller than the beam of our ALMA observations, and therefore, is not detectable in our images. This means that with ALMA observations of the ring only, one would not be able to disentangle a projected circular ring from a ring of eccentricity 0.2.

On the other hand, the deprojected offset between the star and the ring centre of symmetry is $ae$, that is, it is expected to be one order of magnitude higher than the difference between the semimajor and semiminor axis. Indeed, the projected offset as measured in \textit{HST} observations is $\sim 0".8$. This is precisely the size of our ALMA beam, and therefore, the eccentricity of the ring can be retrieved by assessing the position of the star, and measuring its offset from the centre of symmetry of a circular ring model. Note that we have found that the star is not strictly at the center of our image, and therefore, a correction will be applied to the offset found in our MCMC run to retrieve the offset from the star. 

\begin{figure*}
\gridline{\fig{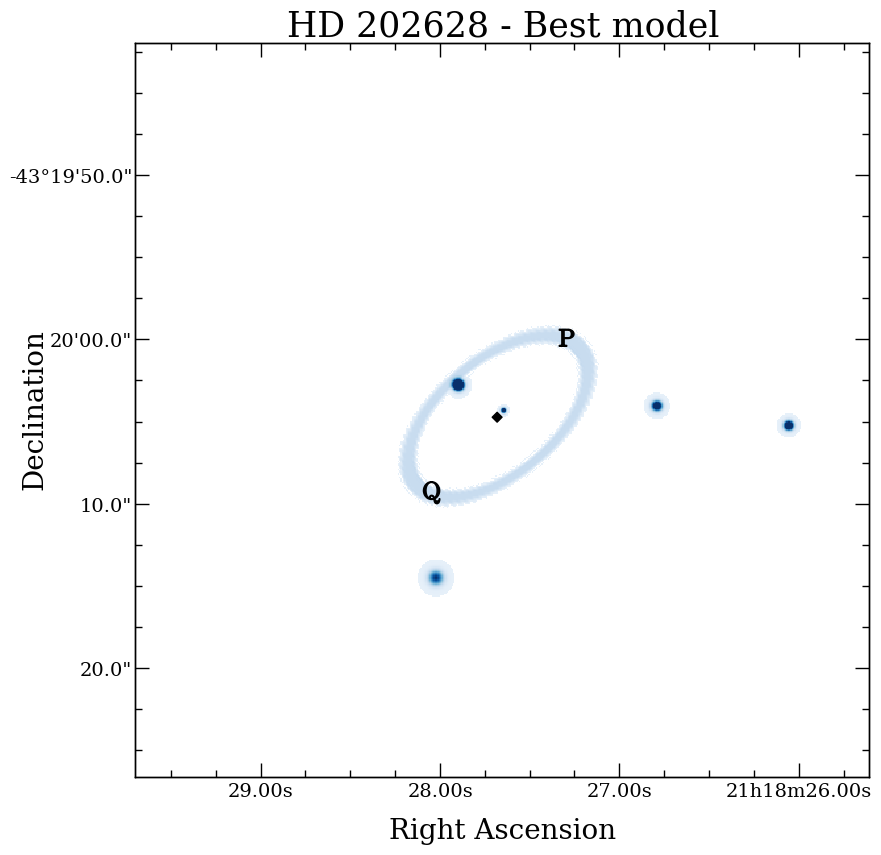}{0.45\textwidth}{(a)}
\fig{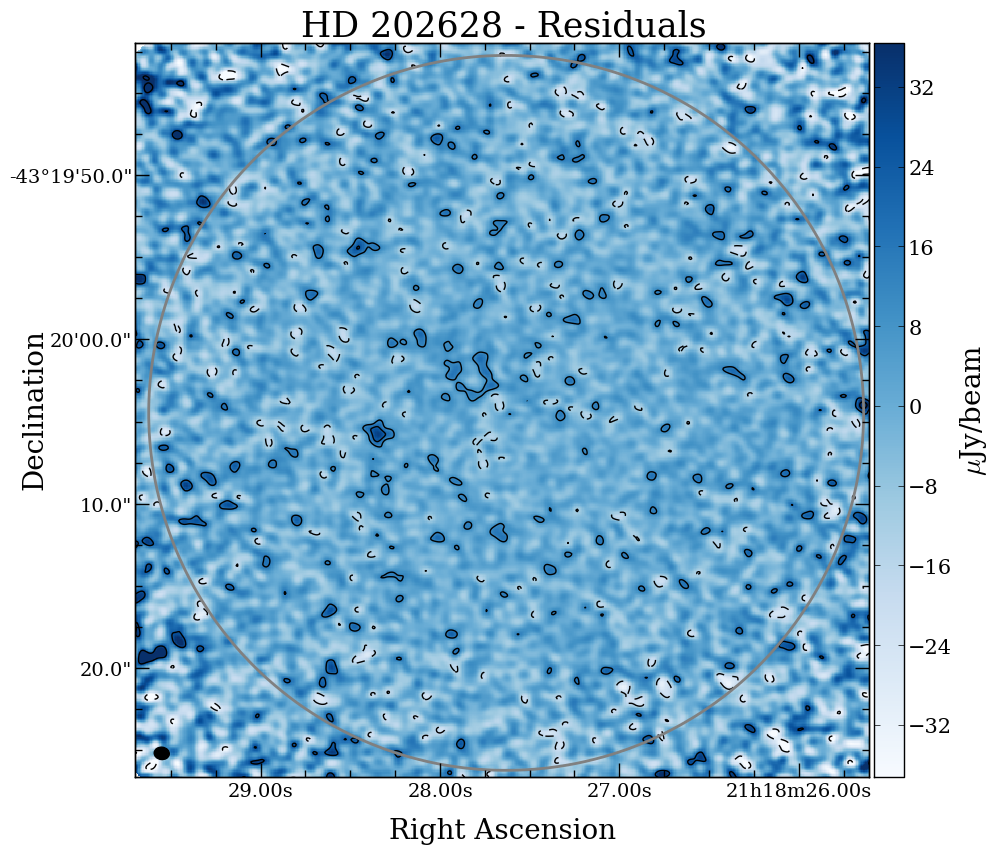}{0.51\textwidth}{(b)}
 }
\caption{(a) Best fit model image, and (b) Residuals of our best fit model with $\pm2\,\mathrm{and}\,4\sigma$ contours, with $\sigma=5.2\,\mu\mathrm{Jy/beam}^{-1}$. The P and Q letters denote the pericenter and apocenter of the ring, respectively, while the diamond shows the ring center of symmetry.}
\label{fig:model}
\end{figure*}

The other free parameters of the disk are: its mass, inner radius, outer radius, inclination, and position angle, for a total of 7 free parameters. It is assumed to have a constant surface density. While the disk geometry derived via MCMC is dependent on accurate modeling of the surface brightness distribution, the ring is only barely resolved in width, and variations of the surface density are not expected to be significant in this case, hence our choice of a constant surface density.
Since the emission from the source S1 overlaps that of the ring, we include a Gaussian source in our model, allowing for its peak brightness, width, and offset from the image center in Right Ascension and Declination to vary. We included as well the sources S2 to S4 in our model as Gaussian emissions. The total number of free parameters in our model is therefore 23. In the frame of a MCMC fitting procedure, these 23 parameters are allowed to vary, and the corresponding model disk emission for each parameter set tested was Fourier transformed into visibilities for comparison with our ALMA data. We summarize our findings in Table \ref{tab:ring_geom}. In Figure \ref{fig:model}, we show our best fit model, along with the residuals once the model is subtracted from our observations.

\begin{deluxetable}{cc}[tbp]
\tablecaption{Result of the MCMC fit on the ring geometry. Error bars correspond to the 16-84 percentiles values. The mass represents the mass in solids of size comparable to the wavelength of the observations ($\sim 1\,$mm), and is thus a lower limit on the mass of the debris disk. \label{tab:ring_geom}}
\tablewidth{0pt}
\tablehead{
\colhead{Parameter} & \colhead{Value}  
}
\startdata
$\mathrm{M_{disk}\,(M_{\oplus})}$  &  $(1.36\pm0.06)\times10^{-2}$  \\[2pt]
Inner radius (AU)  & $143.1\pm 1.7$  \\[2pt]
Outer radius (AU) & $165.5 \pm 1.4$  \\[2pt]
Inclination $(^\circ)$ & $57.4\pm 0.4$    \\[2pt]
PA $(^\circ)$ & $-50.4_{-0.5}^{+0.4}$    \\[2pt]
Eccentricity & $0.09_{-0.01}^{+0.02}$    \\[2pt]
\enddata
\end{deluxetable}

In order to characterize further the geometry of the ring, and in particular to translate its offset from the star into an intrinsic eccentricity, we use the fact that the center of the ellipse, its focus, and its periastron, remain aligned in the projected ellipse. The ring's true eccentricity is given by the ratio between the deprojected offset $ae$ and the ring deprojected semimajor axis $a$, however, since these two quantities are reduced by the same factor during the projection because they share the same direction, the ring's true eccentricity is thus also equal to the ratio between the projected offset and the projected semimajor axis.

We find that the projected offset of the ring centre of symmetry from the star is $0\arcsec.55_{-0\arcsec.1}^{+0\arcsec.09}$, while the ring's inner edge projected semimajor axis is $5\arcsec.9\pm0\arcsec.1$, which leads the ring's true eccentricity to be $0.09\pm0.02$. In addition, the pericenter is found to have a position angle of $318^\circ \pm 10^\circ$ and the apocenter $138^\circ \pm 10^\circ$.

Note that the value of $0.09\pm0.02$ for the ring eccentricity is half the one found by \citet{Krist2012}, but that the orientation and eccentricity of the 2012 image was derived from purely visual ellipse fits to the inner apparent edge of the ring, which was not very well defined (and still isn't even in our much deeper data).  Similar, independent visual fits to the new, combined \textit{HST} yields an eccentricity of $0.14 \pm 0.02$ with an inclination of $56.7 \pm 1.5 ^{\circ}$ and major axis PA of $-52.9 \pm 1.3 ^{\circ}$. While the inclination and PA of the disk are compatible across both ALMA and \textit{HST} datasets, a small discrepancy remains on the eccentricity, which is likely due to the difficulty in defining the inner edge of the diffuse ring in \textit{HST} observations.

Finally, using our best fit model of the sources, we subtracted them to produce an image of the ring in which the source S1 in particular would be disentangled from the ring. This allows us to determine the total flux of the debris ring, which, as delimited by its $2\sigma$ contour, is found to be $959\pm 96\,\mu\mathrm{Jy}$ (including a 10\% absolute flux uncertainty).

Note that in Figure \ref{fig:HD202628_continuum}, our ALMA observations seem to show an emission enhancement in the SE direction, that is, in the direction of the apocenter. This would be expected from an eccentric debris disk, which is bound to be intrinsically asymmetric in azimuth, with the apocenter region being denser than the pericenter region, leading to a so-called "apocenter-glow" at millimeter wavelengths \citep{Pan2016}. However, while this phenomenon should be expected from an eccentric ring such as the one of HD 202628, our observations lack the sensitivity for us to be formally conclusive. This is confirmed as our best fit model is that of an azimuthally uniform ring, and that no residual remain at apastron. Hence the detection of the apocenter-glow remains qualitative.

\subsubsection{Gas content}\label{sec:ALMA_gas}

Although no clear CO J=2-1 emission is observed in the datacube, we use the spectro-spatial filtering technique of \citep{Matra2015,Matra2017a} to boost the SNR of any line emission that may be present but diluted over many spatial and spectral resolution elements. In order to do so, we proceed with the assumption that any gas present is co-located with the belt's continuum emission and moving at Keplerian velocity around the central star, as this has been the case in the vast majority of gas-bearing debris disks observed so far.

We begin by shifting the spectra in each of the cube's pixels by the negative of the Keplerian velocity expected at that belt location. To calculate the Keplerian velocity field, we use the best-fit longitude of ascending node, argument of pericenter and inclination to the line of sight inferred from continuum emission, and a stellar mass of 1 M$_{\odot}$ (see Appendix \ref{app:eric}). We try positive and negative inclinations as its sign (or similarly, the rotation direction) is unknown, and this determines whether either the NW or the SE ansa is approaching Earth.

Then, we integrate emission spatially over a $\sim1\arcsec$-wide elliptical mask covering the region where the ring's continuum emission is detected at a $>2\sigma$ level. This leads to a spectro-spatially filtered 1D spectrum for each sign of the inclination, with RMS noise levels of 2.3 mJy for the native channel size of 0.63 km/s. No significant emission is seen in either spectra at the radial velocity of the central star \citep[12.0$\pm$0.2 km/s in the barycentric frame,][]{Collaboration2018}.
We therefore set a $3\sigma$ upper limit on the integrated CO J=2-1 line flux within the HD202628 belt of 12 mJy km/s, assuming the spectro-spatially filtered line is unresolved \citep[as would be expected, see e.g.][]{Matra2017a}, and adding a 10\% flux calibration uncertainty in quadrature.

To obtain an upper limit on the total CO mass, we assume optically thin emission and calculate CO rotational excitation using the non-LTE (Local Thermodynamic Equilibrium) code of \citep{Matra2015}, including the effect of UV/IR fluorescence \citep{Matra2018}. To calculate excitation of electronic (UV) and vibrational (IR) levels leading to fluorescent excitation, we use the interstellar radiation field (ISRF) from \citet{Draine1978} with the long-wavelength addition of \citet{Dishoeck2006}, superimposed to a PHOENIX photospheric model fitted to observed stellar fluxes (T$_{\rm eff}$=5780 K, log(g)=3.6, [M/H]$=-0.1$), and rescaled to represent the flux received at the ring location. 

We explore the entire parameter space of density of collisional partners \citep[here assumed to be electrons, but note that this does not affect the result, see ][]{Matra2015}, and kinetic temperatures between 10 and 250 K. This leads to a range of $3\sigma$ CO mass upper limits between $1.4-26\times10^{-7}$ M$_{\oplus}$. This  is four to five orders of magnitude smaller than the disk dust mass (see Table \ref{tab:ring_geom}). Consequently, the condition for a narrow ring to be formed as a result of interactions between solids and gas stated by \citet{Lyra2013}, namely a dust-to-gas mass ratio smaller than 1, is not fulfilled, and we can therefore discard this scenario for the debris ring of HD 202628.

A basic calculation of the column density assuming the ring is circular, face-on and uniform in azimuth leads to a value of $4\times10^{12}$ cm$^{-2}$, and allows us to set an upper limit on the optical depth of 0.04 \citep[Eq. 3 of][]{Matra2017}. This confirms that any CO emission that may be present below the detection limits is optically thin at mm wavelengths. 

Such low levels of CO, even assuming the presence of H$_2$ with a low CO/H$_2$ abundance ratio of 10$^{-6}$, cannot be shielded over the lifetime of the system against photodissociation from ISRF UV photons (which dominate over the star at the ring's location), using shielding factors from \citet{Visser2009}. That means that any CO that may be present will be photodissociated in $\sim120$ years, and cannot be primordial in origin. The only possibility would then be continuous replenishment through exocometary gas release, as observed in a number debris disks so far. 

In this scenario, assuming steady state CO production and destruction through the collisional cascade allows us to estimate the mass fraction of CO (+CO$_2$, given CO$_2$ could also be rapidly photodissociated, contributing to the observed CO) ice in the belt's exocomets \citep[see][for details]{Matra2017a}. For HD202628, given its belt average radius of 155 AU and width of 21.5 AU, for a stellar luminosity of 1 L$_{\odot}$ and mass of 1 M$_{\odot}$, and a fractional luminosity of 1.4$\times10^{-4}$, we obtain an upper limit on the CO(+CO$_2$) mass fraction of exocomets of 36-91\%. This is consistent or above values found in other exocometary belts and Solar System comets \citep[see Figure 6 of][]{Matra2017a}, which indicates that it is still possible for exocomets in HD202628's belt to have typical CO ice abundances despite the non-detection presented here.


\subsection{\textit{Herschel}/PACS and SPIRE Observations}\label{sec:hst_res}

\begin{figure*}
\makebox[\textwidth]{\includegraphics[scale=0.6]{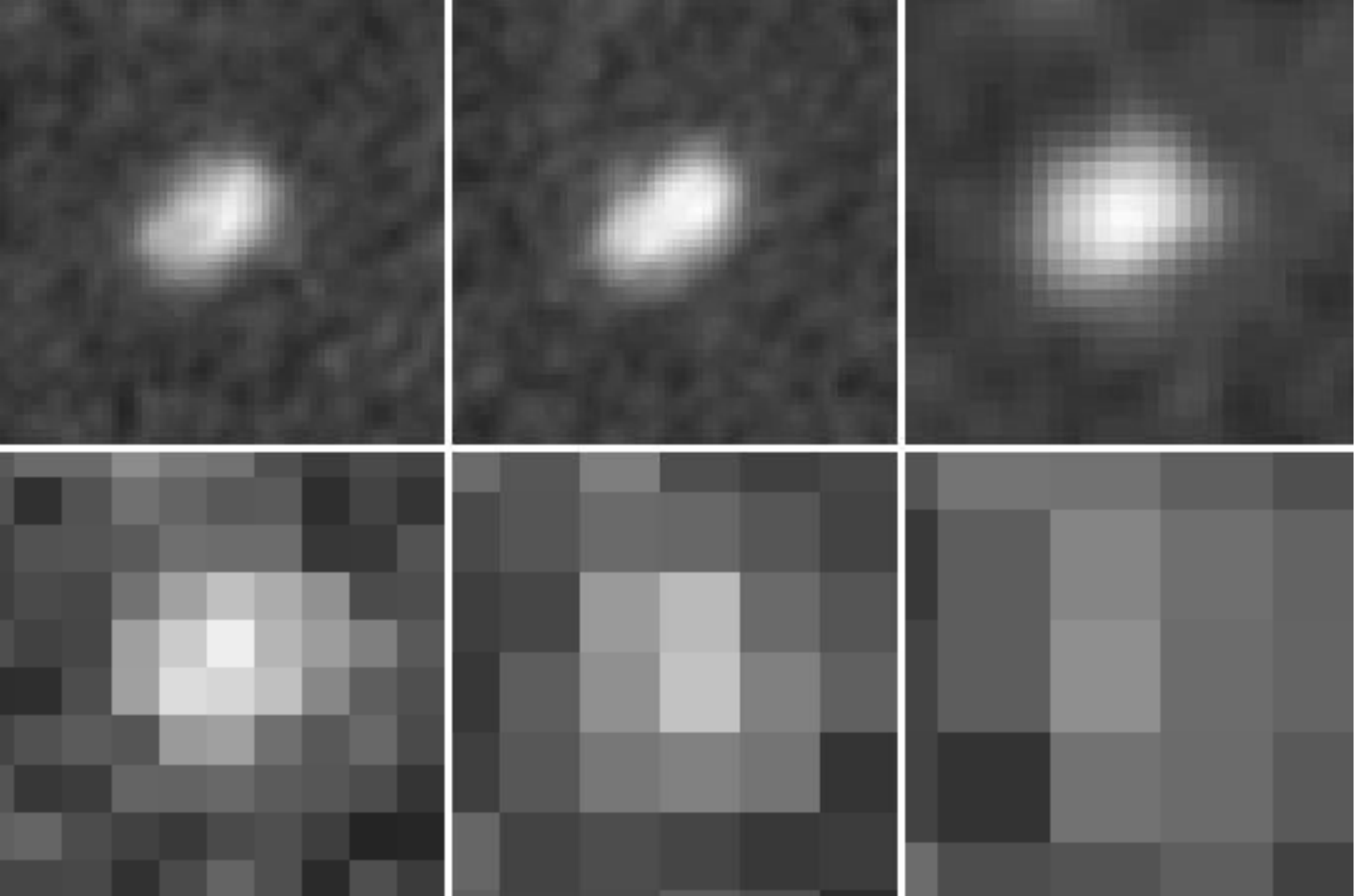}}
\caption{HD 202628 maps as seen in the 70, 100, and 160 $\mu$m channels of \textit{Herschel}/PACS (top row) and the 250, 350, and 500 $\mu$m channels of \textit{Herschel}/SPIRE (bottom row). The pixels have sizes  $1\arcsec.0$, $1\arcsec.0$, $2\arcsec.0$, $6\arcsec$, $10\arcsec$, and $14\arcsec$, respectively, while the corresponding \textit{Herschel}'s resolution is $\sim6$ times these values. The source is detected in all six channels, and clearly extended at the PACS wavelengths.  The field of view of each panel is $55\arcsec$, with North up and East to the left.}
\label{fig:Herschel}
\end{figure*}

The \textit{Herschel} image mosaics are presented in Figure \ref{fig:Herschel}.  
At 70$\mu$m the resolution is sufficient to see the central clearing and higher surface brightness on the ring's West side. 
Gaussian fitting finds source major axis FWHM (Full Width at Half Maximum) values of $13.6\arcsec$, $15.4\arcsec$, and $15.2\arcsec$ at the three PACS wavelengths, after accounting for the instrumental beamsizes reported by \citet{Bocchio2016}. The major axis PA is consistent with higher resolution measurements at other wavelengths. 
At 100 and 160 $\mu$m the source FWHM of $\sim$ 360 AU are consistent with each other and with the size of the ring as seen in scattered light \citep[][, and Section ~\ref{sec:hst_obs} below]{Krist2012,Schneider2016}, while at 70 $\mu$m the ring FWHM of 320 AU is noticably smaller, suggesting that the emission is dominated by the ring inner edge.  

\subsubsection{Pericenter-glow}\label{sec:PG}

\begin{figure}
\centering
\makebox[\columnwidth]{\includegraphics[scale=0.45]{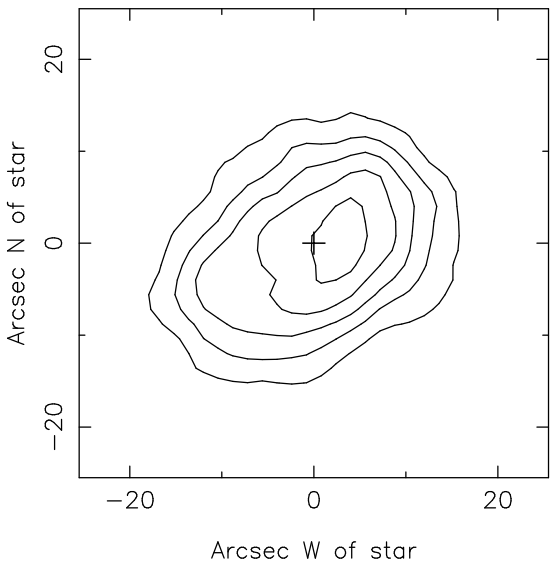}}
\caption{Contour map of the \textit{Herschel}/PACS 70$\mu$m emission from HD 202628. Contours show the $\pm$3, 6, 9,... $\sigma$ significance levels, with $\sigma=29\,\mu\mathrm{Jy.pixel}^{-1}$.}
\label{fig:pacs70_cont}
\end{figure}

A key aspect of the 70 $\mu$m image is its asymmetrically bright emission peak several arcseconds to the West of the star, as shown in Figure \ref{fig:pacs70_cont}. The surface brightness in this region is roughly 30\% brighter than at a comparable distance East of the star. \citet{Krist2012} noted the star's $\sim 0\arcsec.5$ westward displacement from the ring center, a result confirmed by the new ALMA and \textit{HST} results in this paper. 
Asymmetrically bright $\mu$m emission from the ring pericenter is theoretically expected on the short wavelength side of the ring's thermal emission peak \citep[][; "pericenter glow"]{Wyatt1999}, with the effect becoming more pronounced for larger ring eccentricities. It is due to the fact that the pericenter of the ring is closer and thus hotter than the apocenter. At 70 $\mu$m for a ring of eccentricity $\sim 0.1$ such as here, and depending on the dust grains properties, the pericenter is expected to be $20-40\%$ brighter than the apocenter \citep[see Figure 2 of][]{Pan2016}. This is consistent with our findings, and makes HD 202628 only the third debris disk where this has been observed after Fomalhaut \citep{Acke2012} and HR 4796A \citep{Moerchen2011}.  

\subsubsection{Photometry and SED modeling}\label{sec:SED}

Optical and infrared photometry of the HD 202628 system is presented in Table \ref{tab:photometry}.  The values reported for \textit{Herschel}/PACS were measured in apertures of 19.2 arcsec$^2$ at 70 and 100 microns, and of 38.4 arcsec$^2$ at 160 microns, with adjacent background subtracted and uncertainties determined by the 5\% accuracy of the instrument's absolute calibration. Values for \textit{Herschel}/SPIRE were measured in apertures 6, 5, and 3 pixels across, background subtracted, and with uncertainties determined by the observed fluctuations in offset apertures. Extragalactic background confusion could potentially affect the flux densities, with this risk being larger at longer wavelengths. 

\begin{figure}
\centering
\makebox[\columnwidth]{\includegraphics[scale=0.55]{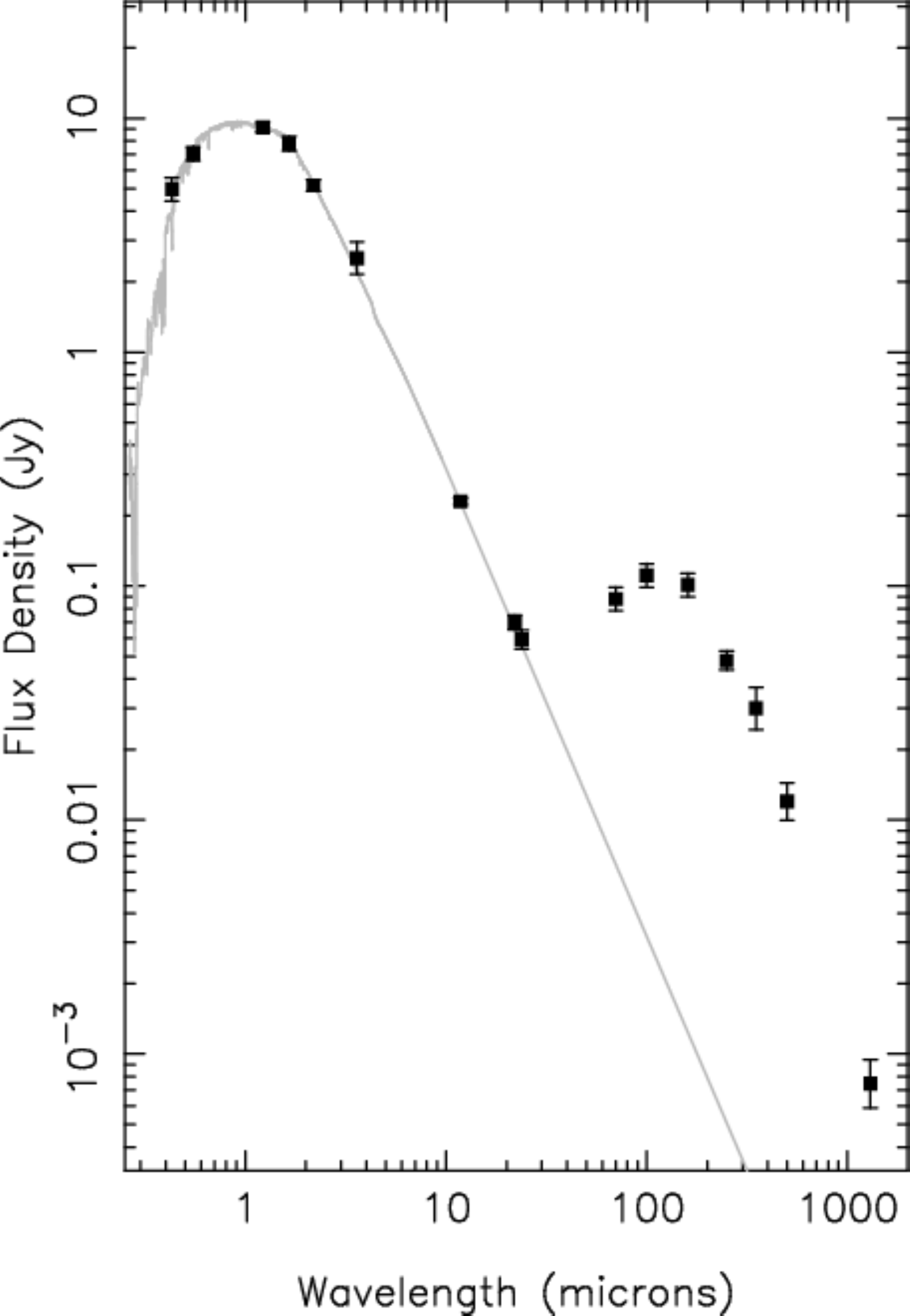}}
\caption{Photospheric model fit and observed flux densities.}
\label{fig:SED}
\end{figure}

To quantify the infrared excess of the disk it is necessary to subtract a model stellar photosphere from the infrared photometry. We used a Kurucz model with T$_{eff}$= 5750, [Fe/H]= 0.0, and log$g= $ 4.5 fit to the 0.43-24.0 $\mu$m datapoints. The observed flux densities and the photospheric model fit are plotted in Figure \ref{fig:SED}.  The Kurucz model fit implies a stellar photospheric emission of 19 $\mu$Jy at $\lambda= 1300 \mu$m, about 20\% less than the value observed with ALMA. This is, however, not surprising, as the few stars for which millimeter emission has been obtained all show fluxes significantly larger than what usual photospheric models predict \citep[see, e.g., ][]{2016A&A...594A.109L}.

The observed spectral index of the debris disk is found to be $2.37\pm0.09$, $2.62\pm0.15$, and $2.64\pm0.20$, using the measured flux at 1.3 mm with ALMA, and the \textit{Herschel}/SPIRE measurements at 250, 350, and 500 microns, respectively. On average, it is thus equal to $2.54 \pm 0.15$. Note that this is compatible with the spectral index of $2.70 \pm 0.17$ that was found for the debris disk of Fomalhaut by \citet{Ricci2012}. Using their Equation (1), and adopting their values for the dust opacity and Planck function spectral indices ($\beta_s=1.8\pm 0.2$ and $\alpha_\mathrm{Pl}=1.84\pm 0.02$), we find that the index $-q$ of the grain size distribution power-law in the debris disk of HD 202628 is $-3.39\pm0.15$, which is in accordance not only with the theoretical Dohnanyi value of -3.5 for a steady state collisional cascade \citep{Dohnanyi1969}, but also with the average value actually observed for debris disks \citep[$<q>=3.36\pm0.02$,][]{MacGregor2016}. On the other hand, it is not compatible with the scenario proposed by \citet{Kenyon2015}, in which a Super-Earth is currently forming at large distance from its host star in this system, as this would require a much steeper grain size distribution, with a power-law index of $q\sim4.5-5.5$.

\begin{figure}
\makebox[\columnwidth]{\includegraphics[scale=0.35,angle=-90]{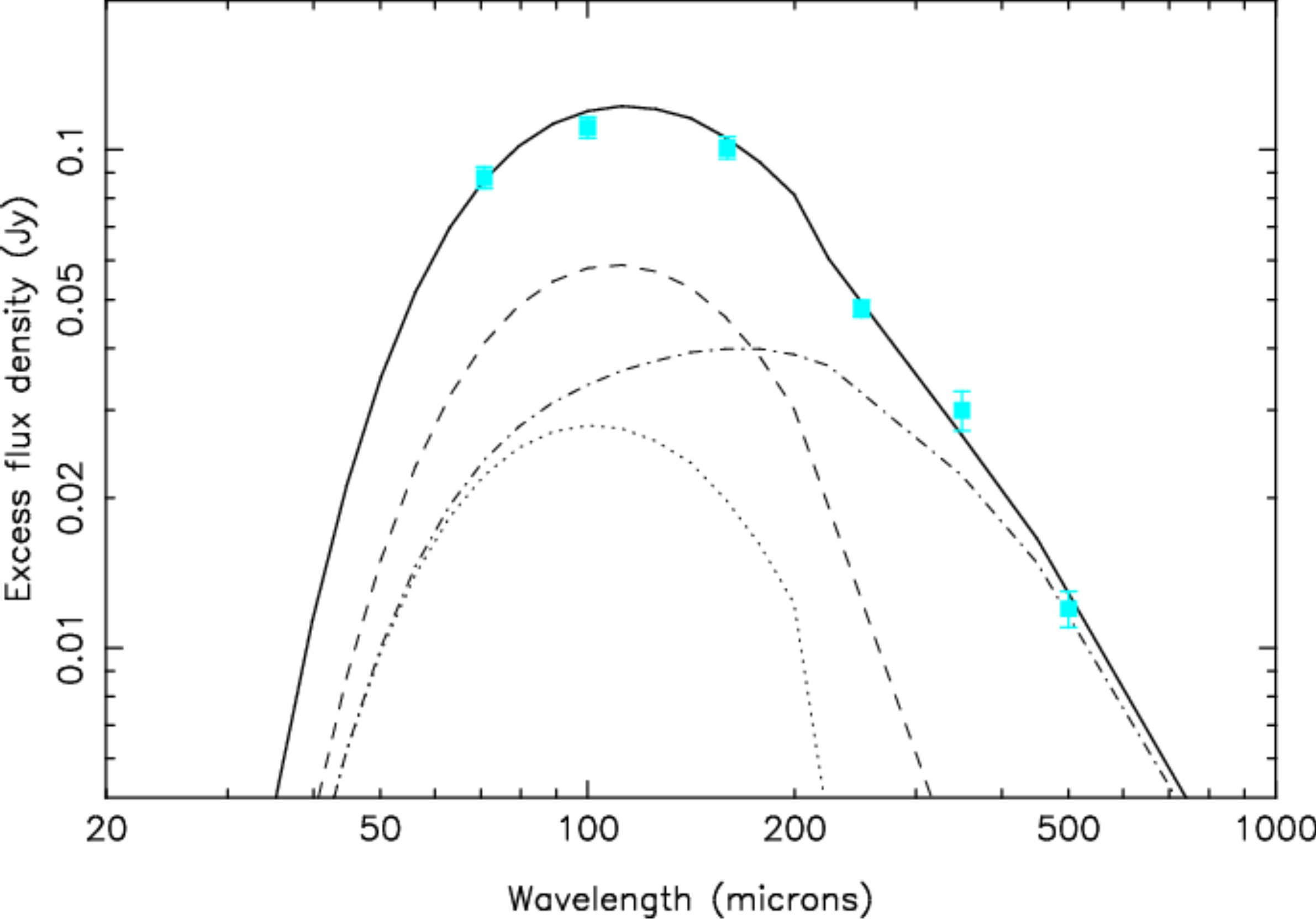}}
\caption{Spectral energy distribution of the infrared excess emission from the HD 202628 debris ring (points) and the best-fit emission model constrained by the three zones derived from the distribution of scattered light.The contributions of the inner, middle, and outer zones are shown by the dotted, dot-dashed, and dashed lines, while the total model emission is shown as a solid line.}
\label{fig:hd202sed}
\end{figure}

The infrared SED can be modeled to further constrain the properties of the circumstellar dust. Following the approach detailed in \citet{Krist2010}, and using the ring size and structure information provided by the \textit{HST} images (see Section \ref{sec:scat-light_model}), a three-zone emission model was constructed with three parameters to describe the dust size distribution: a$_{min}$, a$_{max}$, and the dust size distribution power law spectral index $\gamma$. a$_{min}$ is a free parameter set to the same value in all three zones, while $\gamma$ was fixed to the theoretical Dohnanyi value of -3.5.
While the a$_{min}$ parameter has an important effect on the SED, a$_{max}$ only affects the total millimeter flux, and to a lesser extent, the slope of the SED at the \textit{Herschel}/SPIRE and ALMA wavelengths. Therefore, the most sensitive parameter, a$_{min}$, was iterated first until a reasonably close SED was found, and a$_{max}$ was subsequently set on a second phase. In the central zone corresponding to the parent body ring seen by ALMA, a$_{max}$ was set to 1 mm.  For the inner and outer zones, a$_{max}$ was arbitrarily set to 32 $\mu$m, that is, ten times the minimum grain size, based on the assumption that radiation pressure and Poynting-Robertson (PR) drag cannot effectively transport larger grains away from their formation region in the central zone. Under these assumptions, and using Mie spheres with astronomical silicate composition, the resulting model fit is shown in Figure \ref{fig:hd202sed} along with the excess emission values obtained after subtracting the stellar photospheric model from the flux densities in Table \ref{tab:photometry}.  a$_{min}$ is found to be 3.2 $\mu$m.  As seen in other systems, this value is a few times larger than the theoretical blowout size for a star with 1 $\mathrm{L}_\odot$ \citep{2013MNRAS.428.1263B,2014ApJ...792...65P,2015MNRAS.454.3207P}.  The fractional IR luminosity of best-fit model is $7\times 10^{-5}$, with just over half of this coming from the inner and middle zones.  

\begin{deluxetable*}{cccl}[htbp]
\tablecaption{Photometry of the HD 202628 system. \label{tab:photometry}}
\tablewidth{0pt}
\tablehead{
\colhead{Wavelength} & \colhead{Flux density} & \colhead{Uncertainty} & \colhead{Reference} \\ 
\colhead{($\mu$m)}   & \colhead{(Jy)} & \colhead{(Jy)} &                               \\
}
\startdata
    0.43   &   4.36      &      & Hipparcos \\
    0.55   &   7.50     &      &  Hipparcos \\
    1.22   &   9.15      &  0.18   & 2MASS; \citet{Cutri2003}     \\       
    1.65   &   7.78      &  0.24   & 2MASS; \citet{Cutri2003}     \\
    2.18   &   5.16      &  0.12   & 2MASS; \citet{Cutri2003}     \\
    3.6    &   2.52      &  0.17   & WISE All-Sky catalog  \citep{Cutri2012}      \\
   11.8    &   0.23      &  0.0032 & WISE All-Sky catalog     \citep{Cutri2012}    \\
   22.0    &   0.07      &  0.0022 & WISE All-Sky catalog     \citep{Cutri2012}    \\
   23.8    &   0.059     &  0.0024 & \textit{Spitzer}/MIPS; \citet{Sierchio2014} \\
   70.0    &   0.088     &  0.0044 & \textit{Herschel}/PACS; this work     \\
  100.0    &   0.111     &  0.0055 & \textit{Herschel}/PACS; this work     \\
  160.0    &   0.101     &  0.0051 & \textit{Herschel}/PACS; this work     \\ 
  250.0    &   0.048     &  0.0019 & \textit{Herschel}/SPIRE; this work    \\
  350.0    &   0.030     &  0.0027 & \textit{Herschel}/SPIRE; this work    \\
  500.0    &   0.012     &  0.0010 & \textit{Herschel}/SPIRE; this work    \\
 1300.0    &  0.000959  &  0.000096  & ALMA; this work           \\
\enddata
\end{deluxetable*}


\subsection{New \textit{HST}/STIS observations}\label{sec:herschel_res}

\begin{figure*}
\makebox[\textwidth]{\includegraphics[scale=0.65]{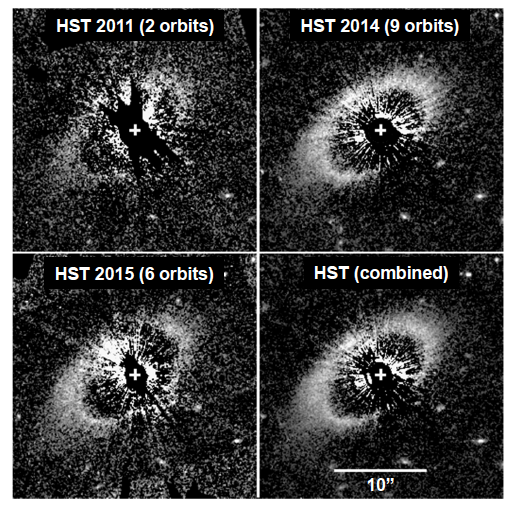}}
\caption{\textit{HST} STIS coronagraph images of the HD 202628 disk derived from ADI post-processing of data from each epoch. In the lower right is the combination. The square-root of the intensity is shown. North is up, and a cross marks the position of the star.}
\label{fig:John_fig1}
\end{figure*}

\begin{figure}
\centering
\includegraphics[scale=0.7]{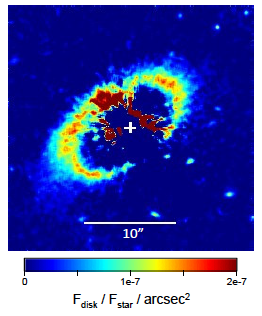}
\caption{\textit{HST} STIS coronagraph image of the HD 202628 disk in units of disk surface brightness relative to the total stellar flux. North is up, and a cross marks the position of the star.}
\label{fig:John_fig2}
\end{figure}

The images in Figure \ref{fig:John_fig1} demonstrate the improvement gained with the additional 7.7 hours of integration time in 2014 and 2015 compared to the original 1.3 hour image from 2011. The inner edge of the ring is better defined and the disk can be traced out further from the star. In the disk ansae the mean signal-to-noise is $\sim 10$, and the mean surface brightness there is $\sim 65$ times less than the stellar PSF's. All of the images show streaks radiating from the star that are subtraction residuals due to PSF mismatches caused by pointing errors and optical aberration variations over time. The disk is likely smooth. Anything interior to the ring may be artifacts as well - there is no infrared excess to suggest interior material that would be visible in these images. Along the ring minor axis the disk suffers the most from subtraction artifacts, and the fainter SW side is especially noisy. A brightening on the NE side of the ring along the minor axis is very likely an artifact. This region is the closest in apparent separation to the star and thus is the most likely to be subject to PSF subtraction artifacts. Indeed, and unsurprisingly, such brightenings appear around this region in the 2014 and 2015 data (which is masked by a spider in 2011), though never at the same position between years or between months within the same year, which is consistent with it being a subtraction artifact.
A calibrated surface brightness map is shown in Figure \ref{fig:John_fig2}. 

There are numerous background galaxies and some point sources visible in the full 50"-wide \textit{HST} image, but there is no correspondence between any of them and sources seen in the ALMA image.

\subsubsection{Deprojection}

\begin{figure*}
\makebox[\textwidth]{\includegraphics[scale=0.7]{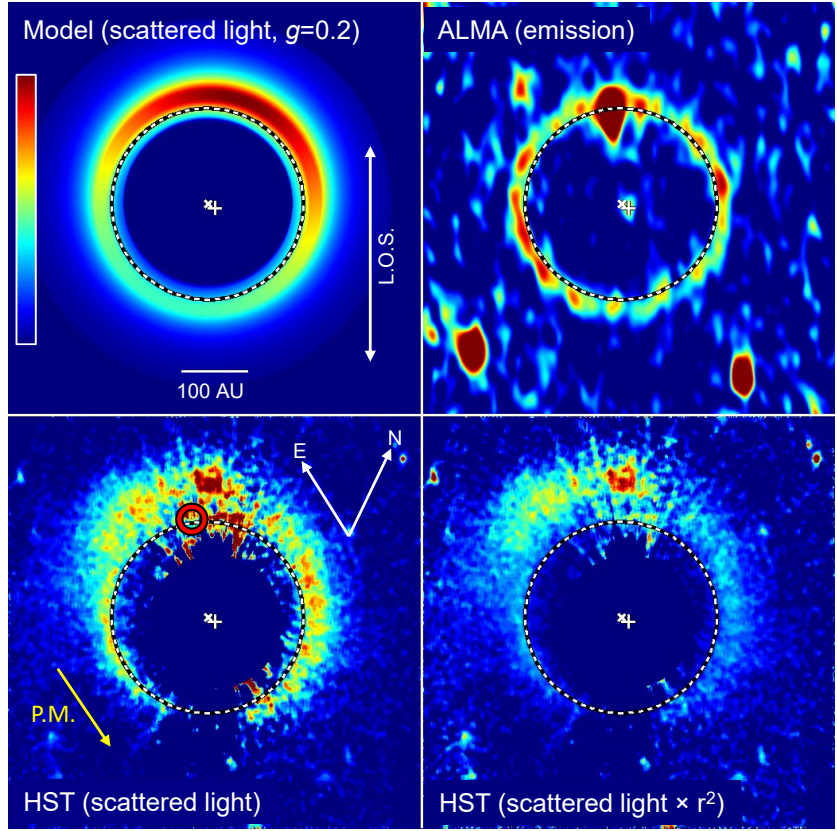}}
\caption{\textit{HST} (PSF-subtracted) and ALMA images of the HD 202628 disk deprojected to a face-on orientation. The horizontal axis is along the line of ascending nodes. The line of sight (L.O.S.) is along the vertical axis. The position of the star is marked by a cross $+$ and the geometric center of the ring by an $\times$. The fitted ellipse to the inner edge of the ALMA ring is represented by a dashed line in each panel. All of the images are displayed with an arbitrary linear intensity scale.  (Top left) Deprojected scattered light model with moderate forward scattering $(g = 0.2)$, assuming that the top side of the ring is closest to the observer. (Top right) ALMA emission image. (Bottom left) Deprojected \textit{HST} scattered light image. The red circle marks where the suspect point source in the ALMA image would be in 2014 due to proper motion of the star if it were a background object, while the arrow represents the deprojected apparent star's proper motion (assuming no out-of-sky motion). (Bottom right) The deprojected \textit{HST} image multiplied by the square of the distance of each pixel from the star, compensating for stellar illumination falloff and representing the relative density distribution of scatters modulated by the scattering phase function.}
\label{fig:John_fig3}
\end{figure*}

Deprojection of the scattered-light emission and accurate knowledge of the geometry of the disk as provided by ALMA can teach us further about the dust grains distribution and properties. Indeed, if the debris disk was a circular ring with a uniform azimuthal density distribution, the observed deprojected scattering pattern would be left-right symmetrical about the line of sight, with the near side of the ring to the observer being brighter than the far side if there is forward scattering. If the ring is not circular but eccentric as is the case here, the side closer to the star will be brighter if the scattering is isotropic (pericenter glow). These effects are illustrated in Figure  \ref{fig:John_fig3} (top left panel), where we display a model of eccentric ring seen face-on, with uniform azimuthal density distribution and moderate forward scattering. The scattering pattern is not exactly symmetrical about the line of sight, but instead is symmetrical about a line that is rotated towards the periastron. Note that as the degree of forward scattering increases, the brightest part of the ring will rotate from periastron towards the line of sight.

Using the inclination $(i)$ and position angle of the ascending node $(\Omega)$ derived from the ALMA image, the ALMA and \textit{HST} images were deprojected via interpolation to a face-on orientation, as shown in Figure  \ref{fig:John_fig3} (top right and bottom left panels, respectively).
We produced as well an additional map where each pixel in the deprojected \textit{HST} scattered light image was multiplied by the square of its physical distance from the star (as derived from the disk geometry seen with ALMA), compensating for the stellar illumination falloff, and shown in Figure  \ref{fig:John_fig3} (bottom right panel). Assuming an optically-and-vertically-thin disk, the result is a map of the relative scatterer density modulated by the scattering phase function of the dust grains. In other words, this map is freed from the effects of stellar illumination, that is, of the pericenter-glow, and provides a better representation of the disk actual density, though not entirely freed from the effects of the scattering properties of the grains. As compared with what would be expected from the disk shown in the model image, the deprojected \textit{HST} image shows that the bulk of the light distribution is instead skewed away from the star and to the left of the line of sight, forming an arc extending radially outward from North to East (counter-clockwise), and which can be explained most easily by a higher relative dust density there (assuming azimuthally-uniform grain properties). This feature is very similar to the one observed in the debris disk of HD 181327, which geometry is precisely the one expected from the distribution of small grains released from a high-mass collisional event \citep{Kral2013,Jackson2014}. This led \citet{Stark2014} to attributed this feature to a massive collision within the debris disk of HD 181327, and hence could be the case for that of HD 202628 as well.

\begin{figure}
\centering
\includegraphics[scale=0.5]{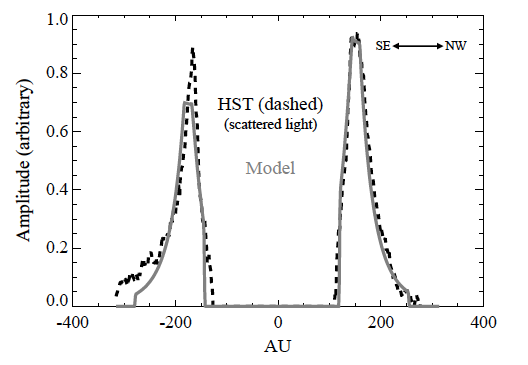}
\caption{Plots of the azimuthal median intensity versus radius from the star measured in $36^\circ$ sectors aligned along the horizontal axis (line of nodes) of the deprojected \textit{HST} STIS coronagraph image of the HD 202628 disk (dashed line) and a corresponding deprojected scattered light model (solid grey line). The intensity scale is arbitrary, and the two plots have been normalized to match on the NW side. The model, which is azimuthally uniform, is brighter to the NW because that is the side closer to the star (the degree of forward scattering is irrelevant since both sides are at the same scattering angle).}
\label{fig:John_fig4}
\end{figure}

While we will focus on the effects of the scattering-phase function in the next section, we can further study the disk azimuthal density distribution using radial profiles.
Figure \ref{fig:John_fig4} shows the surface brightness profile derived by computing at each radius from the star the azimuthal median values in $36^\circ$ sectors oriented along the horizontal axis of the deprojected \textit{HST} image. Recall that this is not the direction of the intrinsic major axis of the ring but rather the axis of the ascending nodes. We precisely chose this direction because the scattering angle along this axis is the same ($90^\circ$), and thus the modulation from the scattering phase function is identical on both sides and has no impact on the relative brightness of each side in the plot. This leaves us with a profile that can be impacted by both the disk eccentricity and its azimuthal density distribution, and freed from the scattering-phase function effects. Since the ring is indeed eccentric, its periastron is expected to be more illuminated and thus brighter in scattered light than the apastron. This effect is shown in Figure \ref{fig:John_fig4} (solid line), and obtained assuming a ring model that is azimuthally uniform. One can see that the observations (dashed line) show instead that the NW and SE sides have comparable brightnesses. This tells us that the ring is in fact not uniform in azimuth, and that its apocenter exhibits an overdensity, as already hinted by the deprojections shown in Figure \ref{fig:John_fig3}.

\begin{figure}
\includegraphics[scale=0.5]{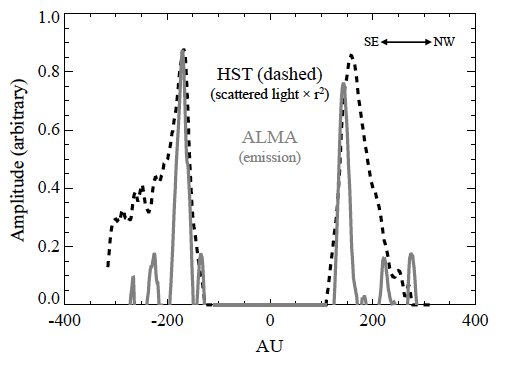}
\caption{Plots of the azimuthal median value versus radius from the star measured in $36^\circ$ sectors aligned along the horizontal axis (line of nodes) of the deprojected HD 202628 \textit{HST} relative density map (dashed line) and the deprojected ALMA image. The vertical scale is arbitrary, and the two plots have been normalized to match on the SE side.}
\label{fig:John_fig5}
\end{figure}

We can take this reasoning a step further by using a similar plot and the deprojected \textit{HST} relative density map, that is, the map which is freed from the pericenter-glow effects. This plot is displayed in Figure \ref{fig:John_fig5} and shows an enhanced density on the SE side, that is, at apastron: this plot shows that the overdensity translates into a ring actually extending further out there, as is expected from eccentric debris disks seen in scattered light, and due to radiation pressure \citep{2016ApJ...827..125L}.
Firstly, small dust grains such as those seen in scattered light are subject to radiation pressure, and are on more eccentric orbits than their parent bodies. Therefore, their distribution extends further out radially than that of the parent grains seen with ALMA. Secondly, an extended apastron in scattered light is due to the fact that the small dust grains are preferentially released on orbits that are apsidally aligned, that is, they are being released in majority at their parent bodies periastron \citep{2016ApJ...827..125L}.

\subsubsection{Scattered light models}\label{sec:scat-light_model}

To qualitatively evaluate various ring properties, such as the scattering phase function and the effect of pericenter glow, we constructed a scattered light model using the measured ring properties. These models were also useful to evaluate the impact of self-subtraction during post-processing, as discussed in Appendix \ref{app:john}. We emphasize the qualitative nature of our model-to-data comparisons presented here: more involved modeling, which would include simultaneously fitting both the \textit{HST} and ALMA data and would account for the apparent azimuthal distribution asymmetry, is left for future studies. A simple single-scattering, three-dimensional model appropriate for optically-thin disks was used with a constant Gaussian vertical distribution with a full-width-half-maximum of 2 AU (the disk is assumed to be so geometrically vertically thin relative to its radial extent that this assumption is largely unimportant in regards to the morphology of the simulated image). Based on the shape of the NW density map profile, the model was composed of three contiguous annular zones, each defined by its inner semi-major axis $(a)$ and a radial density power law:  (1) 132 AU $\leqslant\,a\,\leqslant$ 157 AU, $ r^ {+9.0}$; (2) 157 AU  $\leqslant\,a\,\leqslant$ 172 AU, $ r^ {-2.5}$;  (3) 172 AU $\leqslant\,a\,\leqslant$ 268 AU, $ r^ {-4.5}$. The dust was distributed using the measured eccentricity parameters.

Various scattering phase functions were used: the simple Henyey-Greenstein (H-G) function \citep{Henyey1941}, the \citet{Hong1985} function for zodiacal dust (their $\nu=1$ case), and the \citet{Hedman2015} function for Saturn's G ring. The Hong and G-ring functions are each the sum of three H-G functions with various choices of forward scattering parameters $(g)$ and corresponding weights, and they produce similar results, though the Hong function has slightly more backscatter. See \citet{Hughes2018} for a discussion of the limitations of the single H-G representation for debris disks.

\begin{figure*}
\makebox[\textwidth]{\includegraphics[scale=0.85]{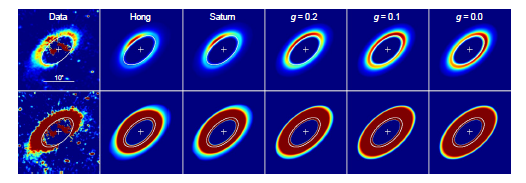}}
\caption{Comparison of the \textit{HST} image of HD 202628 and scattered light models with different scattering phase functions. The top row shows the images with scaling between the minimum and approximately maximum value of each. The bottom row images are truncated at a much lower maximum value to emphasize fainter signal and the isophote of the brightest portions of the disk. The phase functions are Hong zodiacal light, G-ring, and Henyey-Greenstein $(g = 0.2; 0.1; 0.0)$.}
\label{fig:John_fig6}
\end{figure*}

Figure \ref{fig:John_fig6} shows the \textit{HST} data at two different intensity stretches along with the similarly displayed models. Note that in the isotropic scattering $(g = 0)$ case, the ring is brightest to the W, since that is the side closest to the star. Once moderate $(g > 0.2)$ forward scattering is introduced, the brightest portion moves towards the line of sight. The Hong and G-ring phase functions clearly have significant forward scattering (their primary H-G terms have $g = 0.995$ and 0.7, respectively). One must avoid mistaking the bright spot along the NE edge of the data as an indicator of strong forward scattering. It is not located along the line of sight to the star, and it is 4".7 from the star while in the Hong and G-ring models the brightest spots are at 3".9. This further indicates that this feature is likely a PSF subtraction artifact. Even if it were real, it appears more concentrated than the spots in the models. Ignoring that spot, the ring appears much more azimuthally uniform in brightness than in the Hong or G-ring models, suggesting a more isotropic scattering function over the range of observable scattering angles $(g < 0.2)$, as suggested by \citet{Krist2012} and \citet{Schneider2016}.  However, the strong intensity stretch images better match with the more forward-scattering functions, most notably along the presumably far side (SW) of the ring. In the data this side appears truncated when compared to any of the models. The dust distribution is clearly asymmetric, as evidenced by the extension to the SE, which, as mentioned in the previous section, is well reproduced in synthetic images of scattered light emission of eccentric debris disks \citep{2016ApJ...827..125L}.

\section{Constraints on the distant eccentric perturber}\label{sec:planet}

In light of the new constraints for this system, that is, the fact that the ring's eccentricity is found smaller than in previous observations \citep[$e=0.18$][]{Krist2012}, the accurate radial location of its inner edge, as well as the fact that \textit{Gaia} has given new constraints on the star's distance\footnote{Note that the new \textit{Gaia} distance value of 23.8 pc is only $2.5\%$ smaller than the previous \textit{Hipparcos} value of 24.4 pc, and therefore, will only have little impact on the constraints we derive as compared to the accurate knowledge of the ring's geometry.}, we will base our analysis on the previous theoretical modelling work by \citet{Pearce2014} and update the constraints that can thus be set on the eccentric belt-shaping perturber (mass $m_{\mathrm{p}}$, semi-major axis $a_{\mathrm{p}}$, and eccentricity $e_{\mathrm{p}}$). 

We will discuss as well the nature of the source S1, which, with its position just inside the ring, that is, on the expected path of an eccentric perturber, and along with other arguments we develop hereafter, gives us good reason to think it might actually be circumplanetary material surrounding the perturber. We will show how this would allow us to alleviate degeneracies and fully characterize the mass and orbital properties of the belt-shaping perturber.

\subsection{Theoretical constraints}\label{sec:constraints}

Setting precise constraints on the eccentric ring perturber inferred around HD 202628 requires detailed investigations involving complementary theoretical and numerical dynamical modelling, along with the production of synthetic images for comparison with our observations. This type of work has been carried out by \citet{Pearce2014} and \citet{Thilliez2016}, respectively. However, these studies were based on the geometry of the inner edge as given by \textit{HST} in 2012, which was found to be more eccentric than our ALMA and deeper \textit{HST} observations show, and for which the constraints on the radial distance of the inner edge of the star were not as precise as our ALMA observations reveal. Moreover, these studies relied on the distance of the star as found by \textit{Hipparcos} \citep{Leeuwen2007}, whereas \textit{Gaia} has given new constraints on the distance of the system that slightly modify the distance of the inner edge of the disk in AU, and therefore, the constraints derived on the perturber's semimajor axis. Finally, these constraints were derived considering an age of 2.3 Gyr for the system, whereas new estimates give an age of 1.1 Gyr (see Appendix \ref{app:eric}). While extensive parametric exploration and N-body simulations are beyond the scope of the present paper, we nevertheless adopt the theoretical framework of \citet{Pearce2014}, which we summarize hereafter, in order to update the theoretical constraints that can be set on the eccentric belt-shaping perturber at play in the system of HD 202628.

\begin{figure*}[!htbp]
\makebox[\textwidth]{\includegraphics[width=0.5\textwidth,height=0.5\textwidth]{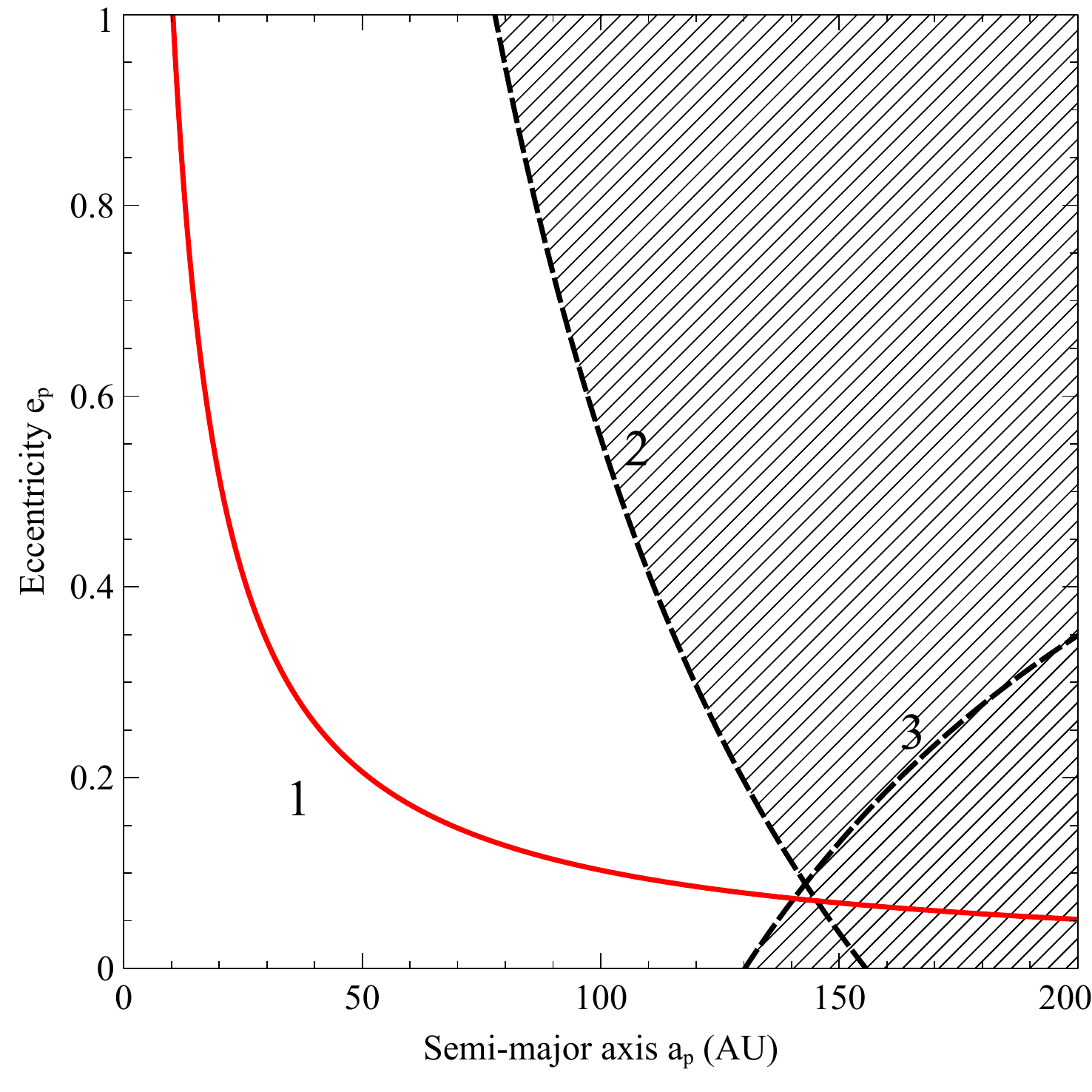}
\includegraphics[width=0.5\textwidth,height=0.5\textwidth]{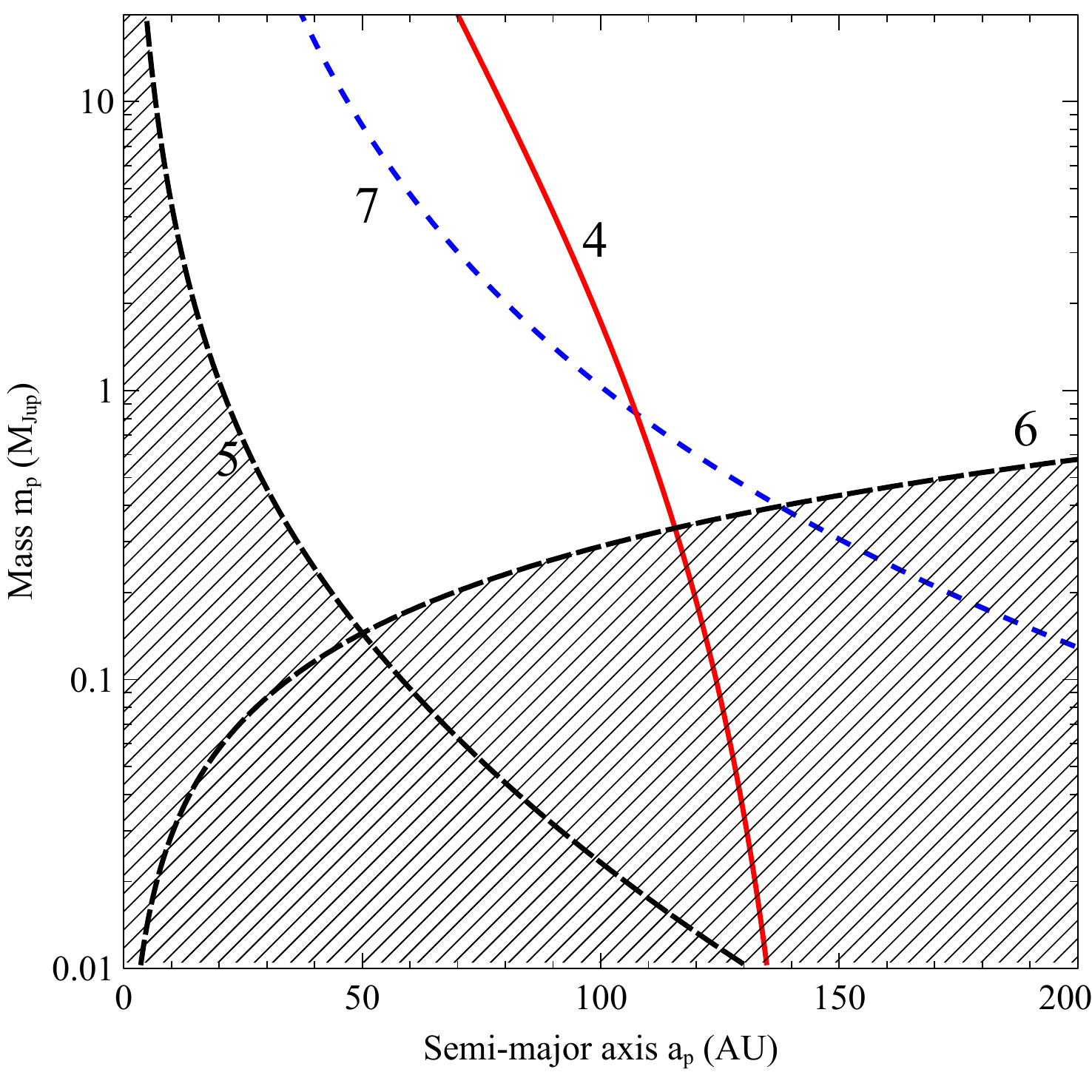}}
\caption{\emph{Left panel:} (1) Forced eccentricity induced on a planetesimal located at the inner edge of the debris ring of HD 202628 as derived from our ALMA observations, that is with semi-major axis $a_{\mathrm{edge}}=143.1\,$AU, as a function of the planet semi-major axis $a_{\mathrm{p}}$ and eccentricity $e_{\mathrm{p}}$. The contour corresponds to the observed eccentricity of $e_{\mathrm{edge}}=0.09$ at the disk inner edge, and thus predicts the combination of parameters $(a_{\mathrm{p}},e_{\mathrm{p}})$ for a planet to induce the observed eccentricity at the disk inner edge.
The two other curves are constraints retrieved from the necessity of the planet orbit to fit within the disk inner edge : (2) the apastron $Q_{\mathrm{p}}=a_{\mathrm{p}}(1+e_{\mathrm{p}})$ of the planet must be smaller than the apastron $Q_{\mathrm{edge}}=a_{\mathrm{edge}}(1+e_{\mathrm{edge}})$ of the disk inner edge, and (3) the periastron $q_{\mathrm{p}}=a_{\mathrm{p}}(1-e_{\mathrm{p}})$ of the planet must be smaller than the periastron $q_{\mathrm{edge}}=a_{\mathrm{edge}}(1-e_{\mathrm{edge}})$ of the disk inner edge.
\emph{Right panel:} (4) Constraint on the mass for the planet to shape the disk inner edge at its observed location, derived from Eq.~(\ref{crit_apo}) and (\ref{rhill_apo}).In addition, the mass of the perturber has to be such that the diffusion timescale, defined by Eq.~(\ref{eq:diff_timescale}), and the secular precession timescale, defined by Eq.~(\ref{eq:secular_timescale}), should be ten times larger than the age of the system. The curve (7) show the couples $(a_{\mathrm{p}},m_{\mathrm{p}})$ that correspond to a Hill radius compatible with the half extent of the source S1 (see Eq. \ref{eq:hill_rad} and Section \ref{sec:circumplanetary}).
}
\label{fig:constraints}
\end{figure*}

Constraints can be set on the perturber's mass $m_{\mathrm{p}}$, semimajor axis $a_{\mathrm{p}}$, and eccentricity $e_{\mathrm{p}}$, relying on secular Laplace-Lagrange theory of perturbations, and considering the following: 

\paragraph{The perturber forces the eccentricity of the debris ring.} The eccentricity $e_{\mathrm{f}}$ forced onto constituants of the debris ring of semimajor axis $a$ reads:
\begin{equation}\label{eq:forced_e}
e_{\mathrm{f}} \simeq \frac{5}{4} \alpha e_{\mathrm{p}} \qquad,
\end{equation}
\citep[Eq. (1) of]{Pearce2014}, where $\alpha=a_{\mathrm{p}}/a$.
From Eq.(\ref{eq:forced_e}), and considering that the inner edge of the ring, with semimajor axis $a=143.1\,$AU, has an eccentricity forced to $e_{\mathrm{f}}=0.09$, one can then derive sets of planetary semi-major axis and eccentricity $(a_{\mathrm{p}},e_{\mathrm{p}})$ that will induce the observed eccentricity at the disk inner edge, as shown  in the right panel of Figure~\ref{fig:constraints}.

\paragraph{The perturber shapes the ring inner edge.} The zone around a planet where mean-motion resonances overlap is called the chaotic zone. Small bodies in this region suffer close encounters, which results in a clearing of this zone. When modelling the shaping of a debris ring inner edge by a perturbing planet, this inner edge lies at the outer boundary of the planet's chaotic zone. 
The width of a chaotic zone depends primarily on the mass and semi-major axis of the planet which is responsible for it, and therefore, one can derive constraints on these two parameters as a function of the inner edge semi-major axis.  
There are different formulae that can be used to derive these constraints, based on previous works, however, as pointed out by \citet{Pearce2014}, most of these results hold for low planet eccentricities, and therefore, they established another criterion which we will use here. This criterion, which was validated numerically, is that the apastron $Q_{\mathrm{edge}}$ of the disk inner edge should lie at approximately five times the Hill radius of the planet at apastron $R_{\mathrm{H,Q}}$ from the planet's apastron $Q_{\mathrm{p}}$, which reads :
\begin{equation}\label{crit_apo}
Q_{\mathrm{edge}}\approx Q_{\mathrm{p}}+5R_{\mathrm{H,Q}} \qquad,
\end{equation}
where $R_{\mathrm{H,Q}}$ is defined by :
\begin{equation}\label{rhill_apo}
R_{\mathrm{H,Q}} \approx a_{\mathrm{p}}(1+e_{\mathrm{p}})\left[ \frac{M_{\mathrm{p}}}{(3-e_{\mathrm{p}})M_{\star}} \right]^{\frac{1}{3}} \qquad.
\end{equation}
For more details, see Eqs.(9) and (10), as well as Appendix B of \citet{Pearce2014}.
As displayed on the left panel of Figure~\ref{fig:constraints}, one can derive an additional constraint on the mass $m_{\mathrm{p}}$ the planet should have to create the disk inner edge when having a semi-major axis and eccentricity compatible with forcing the disk inner edge eccentricity to the observed value, by combining Eqs.~(\ref{crit_apo}) and (\ref{rhill_apo}) with Eq.~(\ref{eq:forced_e}). 

\paragraph{The perturber has acted long enough upon the debris disk for the eccentric ring to be fully formed and well defined.} 
When an eccentric perturber starts acting upon a debris disk (which is assumed to be composed of planetesimals that are initially on low eccentricity, unaligned orbits), the appearance of a well defined eccentric ring is preceded by the appearance and disappearance of spirals. 
As this phenomenon results from the orbital precession of the constituants of the debris disk, the characteristic timescale associated with this phenomenon is the secular precession timescale, which reads:

\begin{equation}\label{eq:secular_timescale}
t_\mathrm{sec} \approx 4 T_\mathrm{p} \left(\frac{m_{\mathrm{p}}}{M_\star}\right)^{-1} \alpha^{-5/2} \left[ b^{(1)}_{3/2} (\alpha) \right]^{-1}  \qquad,
\end{equation}

where $T_\mathrm{p}$ is the perturber's orbital period, $M_\star$ is the mass of the central star, and $b^{(1)}_{3/2}$ is a Laplace coefficient \citep[see Eq. (17) of][]{Pearce2014}. 

\paragraph{The perturber has had the time to clear its surroundings for the inner edge to be sharpened.} Therefore, we consider the diffusion timescale, which is the timescale necessary for a planet to form the inner edge by ejecting material in its surroundings. As derived by \citet{Pearce2014}, but first Tremaine (1993), this timescale reads \citep[see Eq. (18) of][]{Pearce2014}.:

\begin{equation}\label{eq:diff_timescale}
t_{\mathrm{diff}}\sim 0.01 T_{\mathrm{p}} \alpha^{\frac{1}{2}}
 \left( \frac{M_{\mathrm{p}}}{M_{\star}} \right)^{-2} \qquad.
 \end{equation}

In their numerical simulations, \citet{Pearce2014} noted that it took ten times the higher timescale between the secular timescale and the diffusion timescale to form an inner edge and give its eccentricity to the disk. As shown in the right panel of Figure~\ref{fig:constraints}, one can set constraints on the minimum mass of the belt-shaping perturber, as below this mass, it would take timescales larger than the age of the system for the planet to shape the edge of the debris disk and spirals to disappear.

Note that we chose here to focus on the most usual assumption of a single inner perturber, as a thorough dynamical modelling is beyond the scope of this paper. Other possibilities are that the debris disk is shaped by an outer eccentric perturber \citep{Faramaz2014}, or, because of its sharp outer edge, and similar to Fomalhaut, by a pair of shepherding planets \citep{Boley2012}.
In addition, the eccentricity derived for the perturber should be considered to be a minimum and possibly larger. Indeed, it was shown in \citet{Faramaz2014} that a debris disk eccentricity tends to relax to smaller (yet non-zero) values than the predicted forced eccentricity when acted upon by an eccentric planet on timescales of several 100 Myr and up to 1 Gyr.


\subsection{Could S1 be circumplanetary material surrounding the belt perturber?}\label{sec:circumplanetary}

While the ALMA field contains several sources that are all marginally resolved, the brightest one, S1, caught our attention. Indeed, its location is consistent with the expected orbit of the belt-shaping perturber, just interior to the circumstellar ring; therefore, if bound to the star, it could be explained by the existence of circumplanetary ring system surrounding the belt-shaping perturber.
 
No obvious point source is detected with \textit{HST} that is consistent with S1. This is unsurprising because if S1 was part of the system, it could not have been seen in these observations, as its offset of $\sim 3".5$ from the star in the NE direction corresponds to a region which suffers from stellar artefacts and is partially masked by the \textit{HST}/STIS coronograph. 
On the other hand, as the star exhibits a high proper motion (242.190 mas/yr in RA and 21.63 mas/yr in Dec) \citep{Collaboration2018}, a background object would have been offset of $\sim 4\arcsec.5$ NE of the star back in May 2011 on \textit{HST} observations, which show no sign of associated visible light emission at this location, thus placing an upper limit of $V > 25$ on a background source at this location. This cannot completely rule out the possibility of the source being a background galaxy, however, as background galaxies seen at sub-millimeter wavelengths do not obviously show an associated visible light emission. 

According to millimeter-counts in Band 6 and using the Schechter function \citep{Carniani2015}, we would expect 2.4 background sources brighter than S3 (the faintest of the sources) within the ACA Half Power Beam Width (HPBW) of diameter $41\arcsec.4$ in Band 6 (the region delimited by the grey line on the images, that shows the 50\% sensitivity level). Therefore, instead of our circumplanetary ring hypothesis, the presence of a total of four sources in our observations may instead mean that HD 202628 is aligned with galaxies that are part of a cluster, which is the case for 10-20\% of galaxies \citep{Tempel2012}.

This source is not detected in \textit{Herschel}/PACS resolved observations at 70 and 100 microns, which means that if it were circumplanetary material, its smallest dust grains would be larger than $\sim 100$ microns. Note that this would be in accordance with what we know about circumplanetary rings in the Solar System, as the smallest particles in Saturn's A ring have been found to be millimeter sized, possibly down to 500 microns accounting for error bars \citep{Becker2016}. 
It may nevertheless possess emission at \textit{Herschel}/SPIRE wavelengths, for which no resolved image has been obtained, but only photometric measurements, which might actually explain the discrepancy between the flux we predicted for the debris ring and the one we observed. Indeed, we based our Band 6 continuum flux density estimates for the debris disk on these \textit{Herschel} photometric measurements. Following the modeling strategy used for the debris disks of HD 207129 \citep{Krist2010} and HD 92945 \citep{Golimowski2011}, the \textit{HST} image was used to define the spatial distribution of the dust and the dust grain properties were adjusted until the model was consistent with the far infrared SED. Extrapolating the \textit{Herschel}/SPIRE 250-500 micron spectral index to 1.3 mm, which was found to be $2.0 \pm 0.5$, as expected from a solid population in collisional cascade, we had estimated a total 1.3 mm continuum flux density of 1.1-2.6 mJy. We therefore expected to observe the circumstellar ring with a SNR of $\sim 5$ if our predicted flux was distributed in the same way as \textit{HST} scattered light, and expected this SNR to be 10 or higher if the circumstellar ring was much narrower at millimeter wavelengths, similarly to what our ALMA Cycle 4 observations reveal. 
However, we find instead that  the circumstellar ring is mostly detected with a SNR of $\sim 3-4$, and that the disk total flux integrated within the $2 \sigma$ contours is found to be $959\pm96\mu\mathrm{Jy}$. Whilst this is still formally compatible with our estimates, it nevertheless lies at their low end. If the compact source possessed emission at SPIRE wavelengths, we extrapolated the circumstellar debris ring flux from overestimates at these wavelengths, and subsequently would have overestimated the circumstellar disk flux in Band 6. Note that with an integrated flux of $285\pm18\mu\mathrm{Jy}$ for S1, the total flux disk+S1 is compatible with the flux we had predicted for the debris disk.

Assuming a temperature of 30 K and a dust opacity of $2\times 10^{-4} \mathrm{cm}^2\mathrm{/g}$, the source's measured flux density would correspond to a circumplanetary ring system containing a minimum\footnote{This mass is that in grains of size comparable to the wavelength of the observations ($\sim 1\,$mm), and hence does not encompass all the grain size range expected in a disk.} mass of  $10^{-8} M_{\odot}$, that is, 1,000 times the mass of Saturn's rings, or about 1/3 of the Earth's Moon. This would not be the first time such a massive circumplanetary ring system has been detected, as a massive (0.4-8, and possibly up to 100 Moon's masses) ring system has been inferred in the 2MASS J14074792-3945427 system through a complex series of eclipses of the host star \citep{Mamajek2012,Kenworthy2015}. 

Whether the source S1 is co-moving or not cannot be ascertained without second epoch observations. However, if it were to be linked to the system, it would permit us to alleviate the degeneracy that exists between the mass, semimajor axis and eccentricity of the perturber, as we would now have a third dynamical constraint to consider in addition with the debris disk forced eccentricity and inner edge location. Indeed, assuming that circumplanetary material fills the perturber's Hill radius, the measurement of this source's extent would allow us to retrieve this quantity. From our observations, the source's extent is on average 0".58 (13.8 AU) in diameter, which would then correspond to 2 Hill radii of the perturber and we can thus write:

\begin{equation}\label{eq:hill_rad}
2 a_p \left( \frac{m_p}{3M_\star} \right)^{1/3} = 13.8 \mathrm{AU}\qquad.
\end{equation}

As this constraint depends on the mass and semimajor axis of the perturber, we display it in the right panel of Figure \ref{fig:constraints} . One can then read the mass and semimajor axis of the perturber at the intersection and using in turn the constraint on the forced eccentricity, we can retrieve the corresponding perturber's eccentricity.
In that case, the perturber would be characterized by $m_{\mathrm{p}}=0.8\,\mathrm{M_{Jup}}$, $a_{\mathrm{p}}=107.5\,\mathrm{AU}$, and $e_{\mathrm{p}}=0.1$. 
Note that these constraints rely upon strong assumptions regarding the extent of the circumplanetary material, and that it could be expected that it does not entirely fill the Hill radius of the planet \citep{1998ApJ...508..707Q,2009MNRAS.397..657A}, and consequently, the mass derived here for the perturber is only a lower mass limit. In addition, we did not take into account the error bars on the extent of the source, nor the fact that the source might be more extended than found by our best fit, as possibly suggested by the 2-sigma residuals remaining around the source's location (see Figure \ref{fig:model}). Our goal here is simply to show how the knowledge of a third constraint alleviates degeneracies between the mass and orbital properties of the perturber which are usually encountered when trying to characterize a perturber from its gravitational imprint on a debris ring. If this source's extent were larger than the 13.8 AU value we plugged in Equation \ref{eq:hill_rad}, this would suggest that the planet bearing circumplanetary material is more massive than the $0.8\,\mathrm{M_{Jup}}$ value found here, which shall hence be considered a lower limit.

The origin of the only giant circumplanetary ring system we know of relatively well, namely that of Saturn, is still subject to debate, and in particular why Saturn harbours one and not the other giant planets. Three scenarios have been advanced to explain Saturn's ring system: i) the tidal mass removal from a satellite migrating inwards towards Saturn \citep{Canup2010}, ii) the tidal disruption of cometary bodies during close encounters with Saturn \citep{Dones1991,Hyodo2017}, and iii) the collisional destruction of a satellite by a passing comet \citep{Charnoz2009,Dubinski2017}. Scenario i) is often preferred to scenarios ii) and iii), as these both rely on a significant cometary flux, and are therefore deemed fairly improbable in the case of Saturn. However, they certainly would be more plausible in the case the circumplanetary ring system has formed around a belt-shaping eccentric planet such as the one inferred around HD 202628. Indeed, when shaping the inner edge of a debris belt and clearing its chaotic zone, this type of perturber is extremely efficient at setting bodies onto cometary orbits that cross its own orbit \citep[see Figure 6 of][]{Faramaz2015}, thus enhancing the probabilities of close approaches. While the majority of material is scattered rapidly, this dynamical phenomenon nevertheless reaches a steady state potentially ongoing over the system's lifetime, especially if one takes into account the fact that material will steadily diffuse into the planet's chaotic zone and re-feed it. Therefore, since HD 202628 is much older than either Fomalhaut and HR 4796, this phenomenon has most likely been ongoing for a much longer time, and there should be a greater probability for the scenarios ii) and iii) to have occured around HD 202628, giving a potential explanation as to why this would be observed in the HD 202628 system, and not in the Fomalhaut and HR 4796 systems.

Another possible scenario could involve the trapping of dust that migrates inwards due to PR drag \citep{Kennedy2015}. However, small grains are more sensitive to PR drag, which would probably lead the perturber to trap small grains as well, which are not observed. Nevertheless, this scenario could still be compatible with the absence of small grains around the perturber, since these small grains cross more quickly and therefore could be trapped less efficiently. They could also tend to be collisionally destroyed more quickly, or might tend to stick together and grow again once bound to the planet.


\section{Summary \& Conclusions}\label{sec:conclu}

In this work, we have presented a global, multi-wavelength view of the debris ring of HD 202628.

We performed high angular resolution $(\sim 0\arcsec.8)$ observations of the debris disk continuum with ALMA at 1.3 mm, and searched for CO gas emission as well.
In these observations, the star was detected with sufficient SNR to provide accurate knowledge of its position, which is crucial to determine the ring's eccentricity. The ring appears narrower than seen with \textit{HST}, with a width of $\sim 1\arcsec$, that is, barely resolved with our $\sim 0\arcsec.8$ beam.  We used a MCMC fit of our observations in the visibility space to determine the ring's geometry. The inner edge, inclination and position angle are compatible with those found in \textit{HST} observations. Combining this with the knowledge of the star's position allowed us to confirm that the ring is intrinsically eccentric, with eccentricity $\sim 0.1$. This is somewhat smaller than what was originally found with \textit{HST} by \citet{Krist2012}.

No gas was detected. This allowed us to place upper limits on the gas content, as well as on the CO mass fraction of parent bodies. This last value is consistent with what was found in other debris belts. This means that CO could still be present at typical abundances without being detected, but the non detection tells us that any CO emission that may be present below the limits is optically thin and that gas has no influence on the debris ring dynamics. 

Although this result remains qualitative, we find as well that the apocenter of the ring appears brighter than its pericenter, which is in accordance with what is expected of eccentric debris disks at these wavelengths and is called "apocenter-glow". This brightness asymmetry due to an overdensity competes with another brightness asymmetry, the pericenter-glow, which is a thermal effect due to the fact that the pericenter is closer to the star and thus hotter. At ALMA wavelengths, it is expected that the apocenter-glow dominates, while the pericenter-glow is instead expected to dominate at \textit{Herschel} wavelengths. This is exactly what is seen in the \textit{Herschel} observations at 70 $\mu$m, with the West side of the disk being brighter than the East side. This confirms the ring eccentric nature and the position of the ring pericenter on the West side of the ring. Finally, another feature that eccentric rings have been predicted to show in scattered light, as a result from the significant influence of radiation pressure on the micron-sized dust grains these observations trace, is a skirt at apastron. Our \textit{HST} observations confirm this, as the ring extends further out on the apastron side than on the periastron side.

The photometric measurements provided by \textit{Herschel} at both PACS and SPIRE wavelengths have allowed us, in combination with the ALMA ring photometry at 1.3mm, to derive the debris disk spectral index and constrain the grain-size distribution power-law index. This has been found to be $\sim 3.4$, which allowed us to discard the possibility that planet formation was currently occuring in this debris disk, as this would require power law index of 4.5-5.5. Further SED modeling allowed us to determine the smallest grain size, 3.2 microns, which is larger than the blowout size, and to determine a fractional IR luminosity of $7\times 10^{-5}$.

Focusing on the most simple scenario where a single planet interior to the ring shapes its edge and forces its eccentricity, we provide new constraints on the mass, semimajor axis and eccentricity of this putative perturber. Without an additional dynamical constraint, these constraints remain degenerate, as a small mass planet close to the ring inner edge would generate the same ring as a more massive and more eccentric perturber orbiting farther in from the ring towards the star. A third constraint that could alleviate this degeneracy would be some knowledge on the extent of the Hill radius of the perturber, and which could be retrieved if circumplanetary material was detected around the perturber. Our ALMA observations show several bright sources, and we considered the possibility that the source S1 could be such material. Under this assumption, we show how precise constraints on the mass, semimajor axis and eccentricity of the perturber would be retrieved. Nevertheless, it cannot be excluded that this source is in fact a background object, which will be ascertained in the future with second epoch observations (Project 2018.1.00455.S, PI: V. Faramaz) to check for co-movement with the star. In addition, these new observations will be combined with those presented here, possibly leading to a more conclusive detection of the apocenter glow phenomenon.

\acknowledgments

We are very grateful for useful discussions with Mickael Bonnefoy and Julien Milli on exoplanets direct imaging, and with Eve Lee on the effects of radiation pressure on features seen in scattered light in debris disks. 
This paper makes use of the following ALMA data: ADS/JAO.ALMA\#2016.1.00515.S. ALMA is a partnership of ESO (representing its member states), NSF (USA) and NINS (Japan), together with NRC (Canada), MOST and ASIAA (Taiwan), and KASI (Republic of Korea), in cooperation with the Republic of Chile. The Joint ALMA Observatory is operated by ESO, AUI/NRAO and NAOJ. The National Radio Astronomy Observatory is a facility of the National Science Foundation operated under cooperative agreement by Associated Universities, Inc. .
VF's postdoctoral fellowship is supported by the Exoplanet Science Initiative at the Jet Propulsion Laboratory, California Inst. of Technology, under a contract with the National Aeronautics and Space Administration.
MB acknowledges support from the Deutsche Forschungsgemeinschaft (DFG) through project Kr 2164/15-1. 
A.~B., J.~C., and J.~O. acknowledge financial support from the ICM (Iniciativa Cient\'ifica Milenio) via the N\'ucleo Milenio de Formaci\'on Planetaria grant. J.~O acknowledges financial support from the Universidad de Valpara\'iso, and from Fondecyt (grant 1180395).
This work has made use of data from the European Space Agency (ESA) mission {\it Gaia} (\url{https://www.cosmos.esa.int/gaia}), processed by the {\it Gaia} Data Processing and Analysis Consortium (DPAC, \url{https://www.cosmos.esa.int/web/gaia/dpac/consortium}). Funding for the DPAC has been provided by national institutions, in particular the institutions participating in the {\it Gaia} Multilateral Agreement.

\vspace{1cm}

Copyright 2019. All rights reserved.

\appendix


\section{Age}\label{app:eric}

HD 202628 (HIP 105184) is a very well-characterized nearby young solar twin. Its stellar parameters are summarized in Table \ref{tab:stell_param}, and its published age estimates are listed in Table \ref{tab:age}. The star is only slightly hotter than the Sun \citep[G1.5V, T$_{\rm eff}$ = 5833\,$\pm$\,6 K][]{Gray2006,Spina2018}, with metallicity statistically identical to solar ([Fe/H] = 0.003\,$\pm$\,0.004) and surface gravity just slightly higher than solar \citep[(log($g$) = 4.51\,$\pm$\,0.01 compared to 4.44 for Sun;][]{Spina2018}.  This combination of slightly higher temperature and surface gravity led \citet{Spina2018} to estimate a young isochronal age (0.6 Gyr; 0.3-1.1 Gyr, 68\%CL) and slightly higher mass (1.050\,$\pm$\,0.006 M$_{\odot}$). 
Other recent isochronal estimates give similarly young ages: 0.4\,$\pm$\,0.4 Gyr \citep{Reddy2017}, 0.604\,$\pm$\,0.445 Gyr \citep{TucciMaia2016}. However, as can be seen in Table \ref{tab:age}, there are some recent estimates which place the age closer to $\sim$3 Gyr \citep{Soto2018,Casagrande2011}, but those estimates have larger uncertainties.

There are multiple other age indicators which are similarly consistent with a young age for HD 202628. Besides the isochronal age, we discuss multiple age indicators for the star: X-ray emission, chromospheric activity, and abundances of Li, Ba, and Y. All of these are indicators are consistent with the star being slightly older than the Hyades cluster \citep[recent estimates are consistent with $\sim$0.7\,$\pm$\,0.1 Gyr;][]{Brandt2015,Martin2018,Gossage2018,Salaris2018} and significantly younger than the Sun \citep[4.567 Gyr;][]{Amelin2010}.  \citet{Krist2012} reviewed the choromospheric and X-ray activity indicators and adopted an age of 2.3\,$\pm$\,1 Gyr.

As Sun-like stars deplete their Li as they age, trends among age-dated solar twins show that HD 202628's high Li abundace (log $\epsilon$(Li)LTE = 2.23\,$\pm$\,0.02) would similarly appear unusual if the star were $>$2 Gyr \citep[see Fig. 4 of ][]{Reddy2017}, but appears to be clearly older than the $\sim$0.7 Gyr-old Hyades \citep[see Fig. 6 of ][]{Sestito2003}.
Recent spectroscopic surveys of solar twins have shown evidence of strong age-dependent trends in s-process elements \citep[e.g.][]{TucciMaia2016,Spina2016,Reddy2017,Spina2018}. The star's high Barium abundance ([Ba/Fe] = +0.24) is typical for young solar twins, and would appear anomalously high if the star were $>$2 Gyr \citep[see Fig. 5 of ][]{Reddy2017}. Through comparison to an abundance-ratio vs. age trend for age-dated solar twins, the [Y/Mg] abundance ratio for HD 202628 (0.152\,$\pm$\,0.015) derived by \citet{TucciMaia2016} is consistent with an age of 0.8 Gyr.

The star's chromospheric activity indicator (log\,R$^{\prime}_{HK}$) has been reported by multiple surveys, with a wide range of values (maximum of -4.61 reported by \citet{Jenkins2006} to minimum of -4.782 reported by \citet{Gray2006}).  Drawing upon the published activity values from \citet{Henry1996}, \citet{Tinney2002}, \citet{Jenkins2006}, \citet{Gray2006}, \citet{Cincunegui2007}, \citet{Murgas2013}, \citet{Reddy2017}, \citet{Meunier2017}, and \citet{Saikia2018}, we adopt a median estimate of log\,R$^{\prime}_{HK}$ = -4.67.  Using the chromospheric activity-age calibration of \citet{Mamajek2008}, this is consistent with an age of 1.7 Gyr, whereas converting the activity to a predicted rotation period (14 days) would predict an age of 1.6 Gyr. The star's soft X-ray emission was detected in the {\it ROSAT} All-Sky Survey \citep{Voges1999}. Based on the recent reduction of the {\it ROSAT} data by \citet[][; 2RXS catalog]{Boller2016}, we estimate an X-ray activity level of log(L$_X$/L$_{bol}$) = -5.14 and X-ray luminosity L$_X$ = 10$^{28.42}$ erg/s. Using the X-ray vs. age calibration of \citet{Mamajek2008}, this corresponds to an age of 1.2 Gyr, whereas converting the X-ray flux to a predicted rotation period (10.2 days) and then to age via gyrochronology, results in an age of 1.0 Gyr. Taking into account the constraints on the age based on the previously discussed abundance data, the isochronal estimate from \citet{Spina2018}, and the coronal and chromospheric activity age estimates, we adopt a final age of 1.1\,$\pm$\,0.4 Gyr.

To improve estimates of the star's luminosity and radius, and verify the star's mass, we take advantage of the star's similiarity to the Sun to derive improved parameters. HD 202628 is only 61\,K hotter than the Sun. While different Bolometric Correction (BC) compilations sometimes have sizeable zero point differences \citep{Torres2010}, the scales \citep[e.g.][]{Flower1996,Bessell1998,Casagrande2008} are in excellent agreement that the star's BC$_V$ value should be $\Delta$BC$_V$ = BC$_V$ - BC$_{V,\odot}$ $\simeq$ 0.01 mag (few mmag uncertainty) relative to that for the solar effective temperature (T$_{eff,\odot}$ = 5772\,K; IAU 2015 nominal value). We write the bolometric magnitude equation relative to the Sun:

\begin{equation}
M_{bol} = M_{bol,\odot} + M_V - M_{V,\odot} + \Delta BC_V
\end{equation}

Adopting the IAU 2015 nominal solar values and bolometric magnitude \citep[IAU 2015 bolometric magnitude scale is calibrated to nominal solar M$_{bol,\odot}$ = 4.74, tied to nominal solar luminosity and 3.828\,$\pm$\,10$^{26}$ W and nominal total solar irradiance $S_{\odot}$ = 1361 W\,m$^{-2}$;][]{Mamajek2015}\footnote{\url{https://www.iau.org/static/resolutions/IAU2015$\_$English.pdf}}, the solar apparent $V$ magnitude from \citet[][$V_{\odot}$ = -26.76\,$\pm$\,0.03]{Torres2010} which corresponds to solar M$_{V,\odot}$ = 4.812\,$\pm$\,0.03, we estimate M$_{bol}$ = 4.794\,$\pm$\,0.03. Remarkably, the uncertainty is completely dominated by that of the solar $V$ magnitude. 
This translates to luminosity log(L/L$_{\odot}$) = -0.022\,$\pm$\,0.012, i.e. 95\,$\pm$\,3\% that of the Sun. Given the well-defined effective temperature from \citet{Spina2018}, this luminosity corresponds to a photospheric radius of 0.951\,$\pm$\,0.013 R$_{\odot}$.  One can derive a spectroscopic estimate of the star's mass based on the precise estimates of surface gravity and radius.  Adopting log($g$) = 4.438 for the Sun, we estimate the mass to be log(M/M$_{\odot}$) = 0.028\,$\pm$\,0.015 or 1.068\,$\pm$\,0.038 M$_{\odot}$. This corresponds well to the recent estimate based on evolutionary tracks by \citet{Spina2018} of 1.050\,$\pm$0.006 M$_{\odot}$, and empirically confirms the star to be slightly more massive than the Sun independent of theoretical evolutionary models.


\section{Impact of disk self-subtraction due to ADI processing in \textit{HST} observations}\label{app:john}

The ADI post-processing method used here to remove the stellar instrumental pattern from the \textit{HST} images has a
major drawback with extended sources like circumstellar disks: the possibility of source self-subtraction. The algorithm
assumes that the stellar PSF pattern on the detector is constant regardless of the telescope orientation. Any variation
in a given pixel between images is then due to an astronomical source moving through that pixel as the telescope is
rotated about the star. The mean signal that rotates with the same orientation as the telescope then constitutes the
derived source image.

For small, isolated sources like an exoplanet or small galaxy and with sufficient roll, no pixel sees the same astronomical 
source more than once, and its signal is fully assigned to the source image. But if the object is sufficiently
extended, like a disk, such that parts of it pass through a given pixel at two or more orientations, then a portion of
that signal will be considered constant and part of the PSF. These over-bright regions of the PSF would then cause
oversubtraction of the PSF, or effectively self-subtraction of the source. This can be minimized by maximizing the
differences between, and number of, orientations. In the case of a uniform, face-on disk, however, the disk will fully
self-subtract itself.

\begin{figure*}
\makebox[\textwidth]{\includegraphics[scale=0.75]{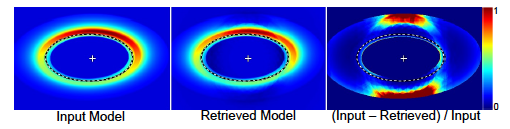}}
\caption{Evaluation of the effect of ADI self-subtraction of the HD 202628 disk in the \textit{HST} 2014 data. \textbf{(Left)} Scattering
model with a H-G phase function and $g = 0.2$ used as input to the ADI algorithm. \textbf{(Middle)} The ADI-retrieved result. \textbf{(Right)}
The relative error between the actual and retrieved model image. All images are shown from minimum to maximum value on a
linear scale.}
\label{fig:John_fig7}
\end{figure*}

To evaluate the potential for self-subtraction, the $g = 0.2$ scattered light model was added to copies of the same
uniform-intensity image (the PSF) at the same nine orientations used in the 2014 observations and sent through the
ADI algorithm. Since this PSF is constant with roll, this experiment solely evaluates the impact of self-subtraction.
The results are shown in Figure \ref{fig:John_fig7} and demonstrate that self-subtraction is minimal $(<15\%)$ over the majority of the
visible disk, including those regions along the line of sight and closest to the star. Significant $(>80\%)$ errors occur in
the outer, farthest reaches of the disk along the line of sight.


\begin{deluxetable*}{lcc}[htbp]
\tablewidth{0pt}
\tablecaption{Stellar Parameters for HD 202628.\label{tab:stell_param}}
\tablehead{
\colhead{Parameter} & \colhead{Value} & \colhead{Reference} 
}
\startdata
$\alpha$ (ICRS, epoch 2015.5) & 319.61505582366 deg &\citet{Collaboration2018}\\ 
$\delta$ (ICRS, epoch 2015.5) &  -43.33455856542 deg & \citet{Collaboration2018}\\ 
Parallax       &  41.9622\,$\pm$\,0.0455 mas & \citet{Collaboration2018}\\ 
Distance       &  23.815\,$\pm$\,0.026 pc & \citet{Bailer-Jones2018}\\ 
Proper motion RA        &  242.190\,$\pm$\,0.068 mas\,yr$^{-1}$ & \citet{Collaboration2018}\\   
Proper motion Dec       &  21.633\,$\pm$\,0.060 mas\,yr$^{-1}$ & \citet{Collaboration2018}\\  
Radial velocity        &  12.071\,$\pm$\,0.0027 km\,s$^{-1}$ & \citet{Collaboration2018}\\  
$V$            & 6.742\,$\pm$\,0.004 mag  & \citet{Mermilliod1990} \\ 
$G$            &  6.5825\,$\pm$\,0.0003 mag & \citet{Collaboration2018}\\
$M_V$          &  4.856\,$\pm$\,0.005 mag & This study\tablenotemark{a} \\
Spec. Type     &  G1.5V & \citet{Gray2006}\\ 
$\mathrm{T_{eff}}$        &  5843\,$\pm$\,6\,K & \citet{Spina2018}\\  
$[\mathrm{Fe/H}]$         & 0.003\,$\pm$\,0.004  & \citet{Spina2018} \\  %
log($g$)       &  4.510\,$\pm$\,0.011 & \citet{Spina2018}\\
U              &  -10.55\,$\pm$\,0.02 km\,s$^{-1}$ & This study\tablenotemark{b} \\
V              &   1.58\,$\pm$\,0.02 km\,s$^{-1}$ & This study\tablenotemark{b} \\
W              &  -28.05\,$\pm$\,0.02 km\,s$^{-1}$ & This study\tablenotemark{b} \\
Mass(tracks)   &  1.050\,$\pm$\,0.006 M$_{\odot}$ &\citet{Spina2018}\\
Mass(spectro.) &  1.068\,$\pm$\,0.038 M$_{\odot}$ & This study\tablenotemark{c}\\
Log(L)          & -0.022\,$\pm$\,0.012 dex & This study\tablenotemark{c}\\  
Radius         &  0.951\,$\pm$\,0.013 R$_{\odot}$ & This study\tablenotemark{c}\\ 
\enddata
\tablenotetext{a}{Calculated using \citet{Mermilliod1990} $V$ magnitude and Gaia DR2 parallax, assuming zero extinction.}
\tablenotetext{b}{Calculated using \citet{Collaboration2018} astrometry and mean published radial velocity.}
\tablenotetext{c}{Spectroscopic mass estimate adopts the surface gravity from \citet{Spina2018}.}
\end{deluxetable*}

\begin{deluxetable*}{lcc}[htbp]
\tablewidth{0pt}
\tablecaption{Age Estimates for HD 202628. \label{tab:age}}
\tablehead{
\colhead{Reference}  & \colhead{Age(Gyr)}    & \colhead{Method}
}
\startdata
\citet{Spina2018}          & 0.6 (0.3-1.1; 68\%CL)     & Isochronal\\
\citet{Soto2018}           & 2.809$^{+1.472}_{-1.505}$ & Isochronal\\
\citet{Santos2016}      & 0.604\,$\pm$\,0.445       & Isochronal\\
\citet{Reddy2017}          & 0.4\,$\pm$\,0.4           & Isochronal\\
\citet{Krist2012}          & 2.3\,$\pm$\,1             & Chromospheric, X-ray\\
\citet{Casagrande2011}     & 3.11 (0.87-6.59; 68\%CL)  & Isochronal(Padova)\\
\citet{Casagrande2011}     & 3.41 (0.96-7.06; 68\%CL)  & Isochronal(BASTI)\\
\citet{Holmberg2009}       & 5.9 (1.6-10.2; 68\%CL)    & Isochronal\\
\citet{Takeda2007}         & 0.64 (0-3.8; 68\%CL)      & Isochronal\\
\citet{Valenti2005}        & 5.0 (1.8-6.8)             & Isochronal\\
\citet{Rocha-Pinto2004}    & 2.27                      & Chromospheric\\
\hline
this study                 & 0.8                       & $[\mathrm{Y/Mg}]$\tablenotemark{a} \\
this study                 & 1.6                       & Chromospheric\tablenotemark{b} \\
this study                 & 1.0                       & X-ray\tablenotemark{c} \\
\hline
{\bf adopted}               & 1.1\,$\pm$\,0.4  & ...\\
\enddata
\tablenotetext{a}{Using $[\mathrm{Y/Mg}]$ abundance and age calibration from \citet{TucciMaia2016}.}
\tablenotetext{b}{Using median log(R$^{\prime}_{HK}$) = -4.67 and calibration from \citet{Mamajek2008}.}
\tablenotetext{c}{Using 2RXS X-ray flux from \citet{Boller2016} and calibration from \citet{Mamajek2008}.} 
\end{deluxetable*}

\bibliography{HD202628_biblio}

\end{document}